\documentclass{hch}
\usepackage{graphicx}
\font\syvec=cmbsy10                        
\font\gkvec=cmmib10                         
\def\bnabla{\hbox{{\syvec\char114}}}       
\def\bxi{\hbox{{\gkvec\char24}}}           
\def\bphi{\hbox{{\gkvec\char30}}}          
\def\spose#1{\hbox to 0pt{#1\hss}}
\def\lta{\mathrel{\spose{\lower 3pt\hbox{$\mathchar"218$}}
     \raise 2.0pt\hbox{$\mathchar"13C$}}}
\def\gta{\mathrel{\spose{\lower 3pt\hbox{$\mathchar"218$}}
     \raise 2.0pt\hbox{$\mathchar"13E$}}}
\begin{document}
\title{Physics Fundamentals of Luminous Accretion Disks Around Black Holes}
\author{O. M. Blaes}
\thanks{This work is supported by NSF grant AST 9970827 and NASA grant
NAG5-7075.}
\address{Department of Physics,\\ University of California,\\ Santa Barbara,
CA 93106, USA}
\runningtitle{Blaes: Accretion Disk Physics}
\maketitle
\begin{abstract}
These lectures provide an overview of the theory of accretion disks with
application to bright sources containing black holes.  I focus on the
fundamental physics of these flows, stressing modern developments and
outstanding questions wherever possible.  After a review of standard
Shakura-Sunyaev based models and their problems and uncertainties, I
describe the basic principles that determine the overall spectral energy
distribution produced by the flow.  I then describe the physics of angular
momentum transport in black hole accretion disks, stressing the important
role of magnetic fields.  Finally, I discuss the physics of radiation
magnetohydrodynamics and how it might affect the overall flow structure
in the innermost regions near the black hole.
\end{abstract}
 
\section{Introduction}

Accretion disk theory was developed in a flurry of activity in the early
1970's in a series of important, seminal papers \cite{pr72,ss73,
nt73,lbp74,le74,pt74,el75,ss76}.
It has met with considerable success in explaining the observations
we see in a number of classes of accretion-powered sources, particularly
cataclysmic variables.  However, it does a rather poor job of predicting
and explaining the observations of black hole sources, both in the context
of X-ray binaries
and active galactic nuclei.  This is particularly disappointing in
view of the fact that much of the early development of the theory was done
with this application in mind.  Accretion
disk theory has always suffered from a severe flaw: angular momentum transport
and energy dissipation in observed flows must be due to nonlinear physics
(``turbulence''), and this problem is swept under the carpet by
parameterizing it away in terms of an anomalous ``viscosity''.  Over the
past thirty years, this sad state of affairs has persisted, although a
major breakthrough was made in 1991 with the discovery that the
magnetorotational instability (MRI) might be a generic source of turbulence
in disks
\cite{bh91,hb91}.  Thanks to this insight and the increasing sophistication
and capabilities of numerical simulations, we are now beginning to
address virtually {\it all} aspects of accretion disk theory from first
principles physics.  The coming decade should see such models applied
to observation, finally getting us to the point where we are able to
test the basic physics, and not fudge factors.  We are therefore living
and working in an exciting time!

In these lectures, I will give an overview of the current state of the
theory.  Although I will be discussing accretion disks in some generality,
I will for the most part concentrate on disks around black holes.
Unless explicitly noted, I will always use Newtonian physics,
even though the flows around black holes necessarily
involve relativistic effects, both special and general.  I do this purely
for the sake of pedagogy.  Relativity is not difficult - in fact, it
is the {\it easy} part of black hole accretion disk physics.  As we will see,
all the really nasty (albeit interesting!) issues lie in the radiation and
plasma physics, and the basic principles involved here are most quickly
understood within a Newtonian framework.  Almost everything I discuss has
been worked out in full general relativity somewhere in the literature, and
I give references to such work wherever possible.

Many good textbooks exist which discuss aspects of the basic physics of
accretion disks, e.g.  \cite{fkr02,krolik99,st83},
and the reader is encouraged to consult them for a more elaborate treatment
of the now-standard ideas.
As I already
mentioned, however, many of these ideas will soon be superseded by much
more sophisticated models based on real physics.  The best overview of
the MRI is the review by Balbus \& Hawley \cite{bh98}, although some advances
have taken place since that article was written.

I will begin in section 2 with a brief overview of standard models based
on the Shakura-Sunyaev $\alpha$-prescription for the anomalous stress that
must exist in these flows, as these are still the models that people use for
comparison to observational data.  I will focus on recent developments in
this area, including the role of advection in ``slim disks'', the role
of torques exerted across the innermost stable circular orbit, and the
vertical transfer of mechanical energy out of the disk into a corona.  In
section 3, I will discuss the calculation of photon spectra from these models.
I will then return to first principles in section 4 by examining the basic
origin of anomalous angular momentum transport in disks, emphasizing the role
of magnetic fields.  Finally, in section 5, I will discuss the role that
radiation magnetohydrodynamics may play in determining the dynamics and
thermodynamics of the innermost regions of the flow, where most of the
observed radiation is thought to originate.
 
\section{Shakura-Sunyaev Based Models}

Most, but not all, models of accretion-powered sources assume that rotation
is an important source of dynamical support in the accretion flow.  Gas
accreting onto a compact object from an orbiting companion star is endowed
with considerable angular momentum arising from the orbital motion of the
binary itself.  The huge dynamic range in radii between the fuel source and
the radius of the central object in young stellar objects (YSO's) and active
galactic nuclei (AGN) also strongly suggests that rotation is important.
If material at large radii has some nonzero angular momentum with respect to
the central object, and if that angular momentum is conserved on the way in,
then the material is likely to become rotationally supported. This is simply
because
centripetal acceleration then depends on distance $R$ from the rotation axis
as $R^{-3}$, steeper than the $R^{-2}$ dependence of gravitational
acceleration.  (This statement neglects general relativistic effects.
Sufficiently close to the event horizon, the gravitational acceleration will
dominate all rotational support and even material with conserved angular
momentum will flow inward all the way to the singularity.)  Direct
observational support
for the existence of rotationally supported accretion flow structures exists
in many images of YSO disks and in the maser disk of the
low luminosity AGN NGC~4258 \cite{miyoshi}.

Angular momentum will generally {\it not} be conserved in the flow, and
the mechanisms of outward angular momentum transport and how they
are connected to the conversion of orbital mechanical energy into other
forms are central to understanding how accretion power works.  Ordinary fluid
viscosity is far too weak to be a significant factor,
and something else must exert torques (either internal
or external, or both) on the flow.

Virtually all models that actually attempt to make contact with the
observations are based on a phenomenological prescription of a
(vertically-averaged) anomalous
internal stress introduced by Shakura \& Sunyaev in 1973 \cite{ss73}:
\begin{equation}
\tau_{R\phi}=\alpha P,
\label{alphastress}
\end{equation}
where $P$ is a vertically-averaged pressure.
Many workers in the field continue to think about this stress as a form of 
``viscosity'', and the stress prescription is often explicitly introduced
as exactly that!  (As we shall see below, such naive thinking
is very dangerous and can be extremely misleading.)  For example, a commonly
used modification of the Shakura-Sunyaev prescription is to assume a
kinematic viscosity of the form
\begin{equation}
\nu\simeq\alpha c_s H\simeq\alpha c_s^2/\Omega_{\rm K},
\end{equation}
where $c_s^2\simeq P/\rho$ is the vertically averaged sound speed (and $\rho$
is a vertically averaged density), $H=H(R)$ is the vertical half-thickness of
the disk, and $\Omega_{\rm K}$ is the angular velocity of test-particle
circular orbits (the ``Keplerian'' angular velocity).
When inserted into the standard viscous form of the stress, this gives
\begin{equation}
\tau_{R\phi}=\rho\nu R\xdrv{\Omega}{R}=\alpha P{R\over\Omega_{\rm K}}
\xdrv{\Omega}{R}.
\label{dodrstress}
\end{equation}
Here $\Omega=\Omega(R)$ is the actual angular velocity in the flow, which
may differ from $\Omega_{\rm K}$ if the flow is not completely geometrically
thin.  If $\Omega\sim\Omega_{\rm K}$, then within factors of order unity,
equation (\ref{dodrstress}) gives the same stress as equation
(\ref{alphastress}).  The only difference is that the stress now depends
explicitly on the shear in the flow, just as an ordinary viscous stress would.
This simple change completely alters the mathematical character of the critical
point problem in stationary advective flows \cite{art01}, but it is not
clear that any of this is real.

Yet another uncertainty that has plagued black hole accretion disk models
with Shakura-Sunyaev stress prescriptions from the very beginning is what
pressure (gas or radiation or...?) to stick into equations (\ref{alphastress})
or (\ref{dodrstress}) when radiation pressure is comparable to or greater
than gas pressure.  This occurs in the central parts of standard disk models
around black holes accreting at anywhere near the Eddington rate (see eqs.
[\ref{prpg1}]-[\ref{prpg2}] below),
right where most of the power is generated.  Moreover, the first
thing that one would try (just inserting the total, gas plus radiation,
pressure) leads to a disk that {\it cannot} be stationary because it
is thermally and ``viscously'' unstable \cite{le74,ss76}.

Let us briefly trace the argument for the radiation pressure driven thermal
instability, which acts on a much faster time scale $\sim1/(\alpha\Omega)$
in geometrically thin disks than the ``viscous'' instability which acts on the
radial flow time scale $\sim R^2/(H^2\alpha\Omega)$.  Assuming that cooling
of the disk proceeds through radiative diffusion, the local emergent flux at
radius $R$ is given by
\begin{equation}
F^-\sim{acT^4\over\tau},
\label{fcool}
\end{equation}
where $a$ is the radiation density constant, $c$ is the speed of light,
$T$ is a measure of the disk interior temperature, and $\tau$ is half
the total vertical optical depth.  The inner parts of black hole accretion
disks have opacities that are generally dominated by electron scattering,
so $\tau\sim\kappa_{\rm T}\Sigma/2$, where $\Sigma$ is the surface density of
the disk, which is constant on these time scales very much less than the
radial flow time scale.  The Thomson opacity $\kappa_{\rm T}$ is also
constant, being
independent of temperature provided there is already sufficient ionization
(and there is), so the optical depth is independent of temperature and
the cooling rate per unit area is $F^-\propto T^4$.  Note that I have made
the usual assumption here that $F^-$ is a real heat flux and does not
involve other forms of energy.

The dissipation rate per unit area is
\begin{equation}
F^+\sim RH\tau_{R\phi}\xdrv{\Omega}{R}.
\label{fheat1}
\end{equation}
Vertical hydrostatic equilibrium implies that the disk half thickness
$H\sim2P/(\Omega_{\rm K}^2\Sigma)$, so that eq. (\ref{fheat1}) becomes
\begin{equation}
F^+\sim{2RP\tau_{R\phi}\over\Omega_{\rm K}^2\Sigma}\xdrv{\Omega}{R}.
\label{fheat2}
\end{equation}
In the radiation pressure dominated inner region, $P\simeq aT^4/3$, so that
equations (\ref{fheat2}) and (\ref{alphastress}) or ({\ref{dodrstress}) imply
that $F^+\propto T^8$!  Hence a perturbative increase in temperature increases
both the local cooling and heating rates, but the heating rate increases much
faster, leading to a thermal runaway.

Such thermal (and ``viscous'') instabilities arising from opacity variations
in hydrogen ionization zones rather than radiation pressure have been applied
with considerable success to understanding the outburst behavior of dwarf
novae (e.g. \cite{osaki96}) and soft X-ray transients (e.g.  \cite{kr98}).  The
reality of these instabilities in radiation-pressure dominated zones has
never been established, however.  It
has been argued \cite{sc81} that the stress should be
proportional to the gas pressure only, in which case
$\tau_{R\phi}\propto\rho T\propto\Sigma^2\Omega_{\rm K}^2T/(4P)\propto T^{-3}$
in the radiation pressure dominated inner zone.  This implies that $F^+$
depends only linearly on temperature, producing a thermally
(and in fact ``viscously'' as well) stable flow.  We will examine this
argument in section 5 below.  In addition, we will look at how radiation
magnetohydrodynamic instabilities might affect thermal instabilities in
accretion flows.

Observationally motivated models of accretion flows generally start with
the assumption that the flow is stationary and axisymmetric about the
rotation axis, even though whatever is responsible for the anomalous
angular momentum transport must almost certainly involve time-dependent,
non-axisymmetric fluctuations.  (Time-dependent models are 
often constructed to model the response of the disk to thermal and ``viscous''
instabilities.)
Another simplifying assumption is that vertically-integrated flow equations 
provide a reasonably accurate description of the behavior of conserved
quantities, even if the flow is not very geometrically thin.
The resulting steady-state conservation laws are those for mass,
\begin{equation}
\dot M=4\pi RH\rho v,
\label{mass}
\end{equation}
radial momentum,
\begin{equation}
\rho v\xdrv{v}{R}=\rho(\Omega^2-\Omega_{\rm K}^2)R-\xdrv{P}{R},
\label{radmomentum}
\end{equation}
angular momentum,
\begin{equation}
\dot{M}\xdrv{\ell}{R}=\xDrv{(4\pi R^2H\tau_{R\phi})}{R},
\label{angmomentum}
\end{equation}
and energy,
\begin{equation}
-4\pi RF_{\rm adv}\equiv
\dot{M}\left[\xdrv{U}{R}+P\xDrv{\left({1\over\rho}\right)}{R}
\right]=4\pi R^2H\tau_{R\phi}\xdrv{\Omega}{R}
+2\pi R(2F^-).
\label{energy}
\end{equation}
Here $\dot M$ is the constant accretion rate through the flow, $R$ is
the distance from the rotation axis, $v(R)$ is the inward radial
flow speed, $\ell(R)$ is the angular momentum per
unit mass, $U(R)$ is the internal energy per unit mass, and $F^-(R)$
is the energy flux leaving the flow on each of the two vertical surfaces.

Note that the anomalous stress $\tau_{R\phi}$ is assumed to enter the
angular momentum
and energy equations in exactly the same way as an ordinary fluid viscous
stress: torques are exerted in a dissipative fashion.  This need not be the
case.  External torques exerted on the disk by a global, ordered magnetic
field in a magnetohydrodynamical (MHD) wind can in principle extract energy
and angular momentum
in precisely the right ratio necessary to allow material to accrete with
no dissipation \cite{bp82}.  The disk can be ice cold and still accreting!  
Because of possibilities like this, I have been careful to call $F^-$ an
energy flux and {\it not} a heat flux.  Even if internal turbulence is
responsible for angular
momentum transport, it is still not clear that all the accretion
power should be converted into heat.  Instead, some of that power may be
converted into bulk kinetic and magnetic energy, and $F^-$ must include
these non-radiative contributions.

Equations (\ref{radmomentum}) and (\ref{energy}) differ from those of standard
geometrically thin accretion disk theory (e.g. \cite{ss73}) by the inclusion
of advective and radial pressure support terms which have received huge
theoretical attention in recent years in the guise of advection dominated
accretion flows (ADAFs, \cite{abram95}, \cite{ny95}).  In the absence of these
effects, one gets the standard equations of thin disk theory, i.e. that
the angular velocity is Keplerian,
\begin{equation}
\Omega(R)=\Omega_{\rm K}(R),
\end{equation}
and local internal dissipation is balanced entirely by local vertical
cooling,
\begin{equation}
F^-=-RH\tau_{R\phi}\xdrv{\Omega_{\rm K}}{R}.
\label{locqbal}
\end{equation}

The angular momentum equation (\ref{angmomentum}) can be integrated from
an inner radius of the flow $R_{\rm in}$ (usually taken to be the innermost
stable circular orbit, or ISCO,  radius for black hole accretion disks) out
to an arbitrary radius $R$ to give
\begin{equation}
\dot{M}(\ell-\ell_{\rm in})=4\pi R^2H\tau_{R\phi}-4\pi R_{\rm in}^2
H_{\rm in}\tau_{R\phi{\rm in}}.
\label{angmomin}
\end{equation}
Combining this with the local energy balance condition (\ref{locqbal})
for geometrically thin disks allows us to write down an expression for
the emergent flux which is independent of the anomalous stress throughout
most of the flow,
\begin{equation}
F^-={3GM\dot{M}\over8\pi R^3}\left[1-\left({R_{\rm in}\over R}\right)^{1/2}
+{4\pi R_{\rm in}^2H_{\rm in}\tau_{R\phi{\rm in}}\over\dot{M}(GMR)^{1/2}}
\right].
\label{fmin1}
\end{equation}
This is the most beautiful result of standard accretion disk theory: that
steady state accretion disks have an emergent flux that is independent
of the details of the anomalous stress.  Of course, accretion flows around
black holes are observed to vary on all sorts of time scales, so
steady state may not be a good assumption.  Moreover, there is still a
dependence on the stress $\tau_{R\phi{\rm in}}$ at the inner edge of the flow.
Until recently, this stress was almost always assumed to vanish near the
ISCO, due to the fact that the inward transonic flow inside this point was
presumed to rapidly decouple from the disk.  This assumption has been
challenged recently \cite{krolik99b,gammie99}, and the physics
of angular momentum transport near the ISCO has since received considerable
theoretical scrutiny.  We will examine
this further in section 4 below, and for now, we will keep the inner stress
in our equations.

Equation (\ref{fmin1}) can be integrated to give the total luminosity of
the disk,
\begin{equation}
L=\int_{R_{\rm in}}^\infty 2\pi R(2F^-){\rm d}R=
{GM\dot{M}\over2R_{\rm in}}+(4\pi R_{\rm in}^2H_{\rm in}
\tau_{R\phi{\rm in}})\Omega_{\rm in}.
\label{luminosity}
\end{equation}
This has a simple physical interpretation.  The first term is the
binding energy per unit mass of material at the inner radius, times the
accretion rate.  The second is the rate at which work is being done on
the disk at the inner radius.  By defining the accretion efficiency in the
usual way, $\eta\equiv L/(\dot{M}c^2)$, with $\eta_{SS}$ being the
efficiency under the standard (Shakura-Sunyaev) assumption of a no-torque
inner boundary condition and $\Delta\eta\equiv\eta-\eta_{\rm SS}$ being
the additional efficiency due to the inner torque, equation (\ref{fmin1})
can be written in a physically appealing way (e.g. \cite{ak2000}),
\begin{equation}
F^-={3GM\dot{M}\over8\pi R^3}\left[1-\left({R_{\rm in}\over R}\right)^{1/2}
+\Delta\eta\left({R_{\rm in}\over r_{\rm g}}\right)\left({R_{\rm in}\over R}
\right)^{1/2}\right].
\label{fmin2}
\end{equation}
Here $r_{\rm g}\equiv GM/c^2$ is the gravitational radius.  (The full
relativistic version of eq. [\ref{fmin2}] may be found in \cite{ak2000}.)
Note that the effects of the inner boundary on the flux decay with radius
as $R^{-7/2}$, only slightly faster than the overall asymptotic decay
of the flux at large radius, $\propto R^{-3}$.

So far we have been neglecting the advective terms in our conservation laws
(\ref{mass})-(\ref{energy}).  If we include them, equation (\ref{fmin1})
becomes
\begin{equation}
F^-=\left(-{\dot{M}\ell\over4\pi R}\xdrv{\Omega}{R}\right)\left(1-{\ell_{\rm in}
\over\ell}+{4\pi R_{\rm in}^2H_{\rm in}\tau_{R\phi{\rm in}}\over\dot{M}\ell}
\right)-F_{\rm adv}.
\label{fminadv}
\end{equation}
Advection, which becomes increasingly important as one moves inward in the
flow, has two effects:  the rotation profile is no longer Keplerian, and the
orbital energy dissipated at any radius is partly advected (lost!) radially
inward.  The latter effect is evident from the last term in equation
(\ref{fminadv}).  The emerging energy flux $F^-$ is less than it would be
if advection was not included.

For luminous accretion disks, advection starts to play a big role when
the overall luminosity starts to get above some fraction ($\sim0.3$ in
standard models) of the Eddington luminosity L$_{\rm Edd}$.  The flow is then
geometrically thicker, becoming a ``slim'' or ``thick'' accretion flow.
Such flows are thermally and ``viscously'' stable \cite{abr81, abr88},
because $F_{\rm adv}$ has a temperature dependence strong enough to
beat the temperature dependence of $F^+$.
The radial and vertical structure of these flows is no longer decoupled,
and the emergent flux $F^-$ now depends on the very uncertain assumptions
that are usually made to solve for the vertical structure.
For a discussion of the fully relativistic equations of advective
accretion flows, see e.g.  \cite{acgl96}, \cite{alp97}, \cite{pa97},
and \cite{bel97}.  Fully relativistic models of optically thick, ``slim''
accretion disks have been constructed in \cite{bel98}.
Models of hyper-Eddington accretion flows onto stellar mass
black holes relevant for gamma-ray burst models have also been constructed
\cite{pwf99}.  By far and away most of the energy dissipated in such flows
is advected into the black hole until the accretion rate exceeds
$\sim10^{-2}$~M$_\odot$~s$^{-1}$, where neutrino losses can cool the gas
on the flow time scale.

Before leaving this section, I should mention the considerable
attention that has been devoted recently to optically thin, radiatively
inefficient ADAF solutions around black holes.
Because we are focusing on luminous accretion flows here, I will have little
to say about these solutions, but it is important to recognize that they
may nevertheless arise over certain ranges of radii even in otherwise
luminous flows.  In particular, it is noteworthy that such flow solutions
exist that are thermally and ``viscously'' stable over much the same range
of accretion rates that geometrically thin, optically thick disk solutions
exist (e.g. \cite{chen95}), and it is not clear why nature should choose
one flow solution over the other.  The presence or absence of an inner ADAF
flow might explain the different spectral states observed in black hole
X-ray binaries \cite{esin97}.  One should bear in mind, however, that all
the models discussed in this section are based on the $\alpha$ stress
prescription, and it is fair to say that the microphysics of optically thin
ADAFs is even more poorly understood than that of optically thick flows
(e.g. \cite{beg88}, \cite{quat99}).

\section{Spectral Formation}

As we saw in the last section, if we neglect the effects of advection, then
the energy flux emerging from each face of a stationary accretion disk can
be written
\begin{equation}
F^-(R)={3GM\dot{M}\over8\pi R^3}{\cal I}_{a/M}(R/r_{\rm g}),
\label{fmingr}
\end{equation}
where ${\cal I}_{a/M}(x)$
is a function which in full general relativistic treatments depends on the
spin parameter $a/M$ of the hole but still approaches unity at large radii.
In standard models with a no-torque
inner boundary condition, ${\cal I}_{a/M}$ also vanishes at the ISCO,
but as we have seen, this need not be the case.  Once again, let me remind
the reader that although $F^-$ is usually assumed to be the emerging
radiation flux from the disk surface, this need not be the case.  It
is actually the total energy flux, and may include contributions from
magnetic and bulk kinetic energy.  We will return to this point shortly,
but for now, let us make the standard
assumption that $F^-$ really does represent an emerging radiation flux.

What, then, is the emerging spectral energy distribution?  An observer here
on earth, viewing the disk at a distance $d$ away at an inclination angle
$i$ to the rotation axis, will measure a spectral energy flux
\begin{equation}
F_\nu={\cos i\over d^2}\int_{R_{\rm in}}^{R_{\rm out}}2\pi RI_\nu(R,i)
{\rm d}R,
\end{equation}
where $I_\nu$ is the angle (limb darkened or brightened!) and radius-dependent
local emergent specific intensity, and we have
introduced a finite outer radius $R_{\rm out}$ to the disk.  (Once again,
we are using Newtonian physics here, and neglecting  e.g. Doppler shifts,
gravitational redshifts, the gravitational bending of light rays, and
the relativistic proper emitting area, but these are all straightforward to
account for using relativistic ``transfer functions'', e.g.
\cite{cun75,speith}.)

The simplest (and therefore most often used!) way of estimating the shape
of the spectrum is to just naively assume that each annulus radiates like an
isotropic local blackbody, i.e. $I_\nu(R,i)=B_\nu[T_{\rm e}(R)]$, where
$B_\nu$ is the Planck function with temperature given by the effective
temperature determined from equation (\ref{fmingr}),
$T_{\rm e}(R)=[F^-(R)/\sigma]^{1/4}$.  If, say, the radial dependence of
the emerging radiation flux can be approximated as a power law,
$F^-(R)\propto R^{-\beta}$, then a simple change of integration
variable to $x\equiv h\nu/(kT)$ implies that the shape of the observed
spectrum is given by
\begin{equation}
F_\nu\propto\nu^{3-8/\beta}\int_{h\nu/(kT_{\rm ein})}^{h\nu/(kT_{\rm eout})}
{x^{8/\beta-1}{\rm d}x\over e^x-1}.
\label{fnubeta}
\end{equation}
As one would expect, at low frequencies satisfying $h\nu\ll kT_{\rm eout}$ we
simply end up with $F_\nu\propto\nu^2$ (the Rayleigh-Jeans portion of the
spectrum emitted by the outermost annulus), while at high frequencies
satisfying $h\nu\gg kT_{\rm ein}$ we have
$F_\nu\propto\nu^3\exp[-h\nu/(kT_{\rm ein})]$ (the Wien portion of the spectrum
emitted by the innermost annulus).  In between, for
$kT_{\rm eout}\ll h\nu\ll kT_{\rm ein}$,
the observed spectrum is a superposition
of blackbodies from many different radii, producing a power-law spectrum
$F_\nu\propto\nu^{3-8/\beta}$.
Now, at large radii, equation (\ref{fmingr})
implies that $\beta=3$, giving the famous result that $F_\nu\propto\nu^{1/3}$
for a multi-temperature blackbody accretion disk.

There are many ways in which this standard result can be modified.  One
possibility is that energy liberated at small radii in the flow is reprocessed
by the disk at larger radii.  By way of
illustration, consider a source of luminosity $L_\star$ located on the
rotation axis at height $H_\star$ above the equatorial plane.  Figure
\ref{hstar} illustrates the geometry of the problem.

\begin{figure}
\centering
\includegraphics[width=9.0cm]{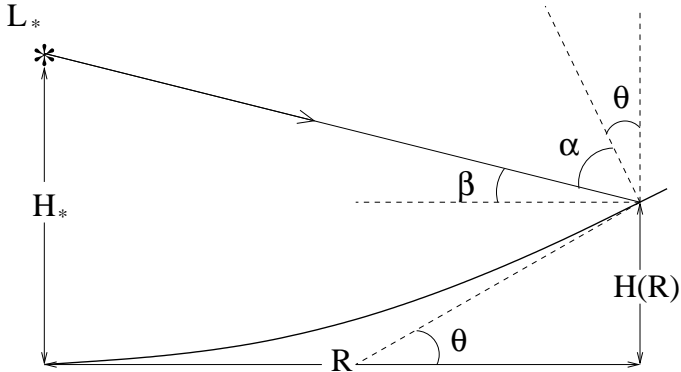}
\caption[]{Illumination geometry for a source on the symmetry axis at
height $H_\star$ above the equatorial plane.  The angle $\alpha$ between the
disk normal and the incoming light ray is simply $\pi/2-(\beta+\theta)$, where
$\theta$ is the angle between the tangent to the local disk surface and
the equatorial plane, and $\beta$ is the angle between the illuminating light
ray and the equatorial plane.  For large $R$, $\beta$ and
$\theta$ are both small angles, with $\beta\simeq(H_\star-H)/R$ and
$\theta\simeq{\rm d}H/{\rm d}R$.}
\label{hstar}
\end{figure}

Neglecting general relativistic effects, in particular light bending, the
irradiating flux on the outer disk at radius $R$ will be given by
\begin{equation}
F_{\rm irr}={L_\star (1-a)\over4\pi R^2}\cos\alpha,
\label{firr1}
\end{equation}
where $a$ is the albedo of the outer disk, averaged over the illuminating
spectrum, and $\alpha$ is the angle between the disk normal and the incoming
light ray.  Taking the illuminating radius $R$ to be much larger than the
local disk height $H$, the source height $H_\star$, and also
${\rm d}H/{\rm d}\ln R$, then a little trigonometry gives
\begin{equation}
F_{\rm irr}={L_\star (1-a)\over4\pi R^2}\left({H\over R}\right)\left(
\xdrv{\ln H}{\ln R} -1+{H_\star\over H}\right).
\label{firr}
\end{equation}

For a source located near the equatorial plane, e.g. the surface of the
inner disk itself, we have $H_\star\ll H(R)$.  Equation (\ref{firr}) then
implies that significant reprocessing will only occur if the outer disk
geometry flares up, with $H/R$ an increasing function of $R$, or is warped
\cite{pringle96}.  In this case, proportionately more power is transferred
to larger radii, and this decreases the value of $\beta$ and reddens the
spectrum.  Standard models of disks work in just this way, with
$H\propto R^{21/20}$ and $R^{9/8}$ for the so-called ``middle'' and ``outer''
regions, respectively.  This gives an irradiating flux approximately
proportional to $R^{-2}$, which must therefore dominate the more rapidly
decaying ($\propto R^{-3}$) local dissipation, and giving
$F_\nu\propto\nu^{-1}$ (e.g. \cite{fkr02}).

For a source located far above the equatorial plane, with $H_\star\gg H(R)$,
equation (\ref{firr}) gives an irradiating flux
$\propto R^{-3}$, producing the same $F_\nu\propto\nu^{1/3}$ spectrum
as a standard disk but with enhanced power.  One way in which this might be
done is by having a geometrically thick, hot corona or ADAF in the inner parts
of the flow.  General relativistic light bending can also produce a similar
effect: an observer located on the outer disk can see the opposite side
of the disk at an effectively high $H_\star$ \cite{ak2000}.  This effect alone
can produce significant additions to the locally dissipated power
in the case of rapidly rotating Kerr black holes \cite{cun76}.

Even without reprocessing, the emergent spectrum can be affected by alterations
in the radial distribution of the local radiative flux $F^-$.  For example,
if an inner torque across the ISCO produces a large increase in the
accretion efficiency, equation (\ref{fmin2}) implies that a substantial
range of inner radii could have $F^-\propto R^{-7/2}$, giving a spectral
energy distribution from equation (\ref{fnubeta}) of the form
$F_\nu\propto\nu^{5/7}$, {\it bluer} than $\nu^{1/3}$!  On the other hand,
gravitational light bending from the much brighter inner disk enhances the
amount of reprocessed flux at larger radii, giving $F_{\rm irr}\propto R^{-3}$
and enforcing a more usual $F_\nu\propto\nu^{1/3}$ spectrum at longer
wavelengths \cite{ak2000}.

Advection also alters the radial distribution of $F^-$, causing it to rise
less steeply toward smaller radii because more of the accretion power is
advected into the black hole rather than being radiated.  This in turn
implies that $F^-$ falls less steeply when moving out in radius, implying
a redder spectrum than $F_\nu\propto\nu^{1/3}$ \cite{szu96}.  In fact, models
of slim disks can give flux distributions as flat as $F^-\propto R^{-2}$,
and equation (\ref{fnubeta}) then gives $F_\nu\propto\nu^{-1}$
\cite{wan99,wat99,min00}.  The global
effect of advection on the spectrum is that the luminosity is less than
you would expect for the given accretion rate, because advection necessarily
reduces the radiative efficiency.

Most models of the vertical structure of accretion disks around black holes
imply that local blackbody emission is likely to be a {\it very} poor
approximation.  The most serious problem with it is the neglect of electron
scattering, which is often far greater than true thermal absorption opacity.
Again, a crude approach to handling this has been used from the very
beginning of accretion disk modeling \cite{ss73}, and that is to use
a so-called local modified blackbody spectrum at each radius,
\begin{equation}
I_\nu={2B_\nu(T)\epsilon_\nu^{1/2}\over1+\epsilon_\nu^{1/2}},
\label{modbb}
\end{equation}
where
\begin{equation}
\epsilon_\nu={\chi_\nu^{\rm th}\over n_e\sigma_{\rm T}+\chi_\nu^{\rm th}}
\end{equation}
is the photon destruction probability, with $\chi_\nu^{\rm th}$ being
the thermal absorption coefficient, $n_e$ the electron number density,
and $\sigma_T$ the Thomson cross-section.

The physics behind equation (\ref{modbb}) lies in the idea that, once created
by thermal emission processes, photons random walk out of the atmosphere.  If
absorption opacity is smaller than electron scattering opacity, this random
walk does not destroy photons generated within an effective optical depth
$\tau_{\rm eff}=(\tau_{\rm abs}\tau_{\rm T})^{1/2}$ of the surface, where
$\tau_{\rm abs}$ is the true absorption optical depth and $\tau_{\rm T}$ is
the Thomson depth.  Assuming LTE and integrating the emission over this
vertical layer immediately gives $I_\nu\sim B_\nu\epsilon_\nu^{1/2}$,
in rough agreement with equation (\ref{modbb}).  Provided the accretion
disk is {\it effectively} thick ($\tau_{\rm eff}>1$ at the frequency of
interest), then the spectrum may be viewed as coming from a depth in
the disk corresponding to $\tau_{\rm eff}=1$.  As shown in Figure \ref{taum1},
the innermost regions of some black hole accretion disk models are not even
effectively thick, however, and the observed spectrum is formed throughout the
entire geometrical thickness of the disk.  In such cases the spectrum
will be directly sensitive to the assumptions that went into
creating the entire vertical structure.  Even when $\tau_{\rm eff}>1$, the
spectrum will depend on the ambient densities and temperatures at the
$\tau_{\rm eff}=1$ effective photosphere, and these will in turn depend on
the vertical structure of the disk model.

\begin{figure}
\centering
\includegraphics[width=9.truecm]{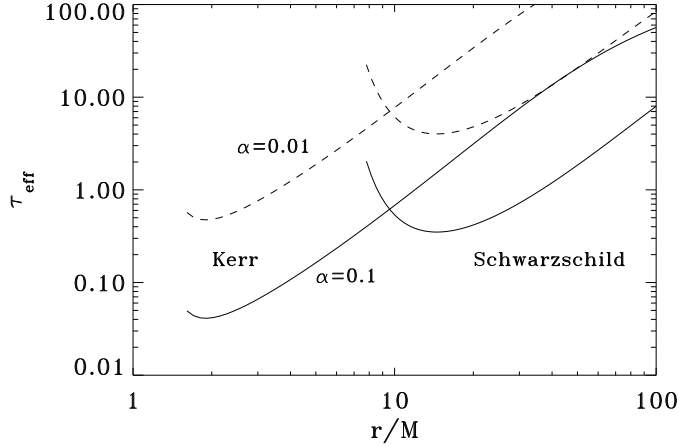}
\caption[]{The midplane effective optical depth as a function of radius in
relativistic accretion disk models around a ten solar mass black hole with
$L=0.3L_{\rm Edd}$ and $\tau_{R\phi}=\alpha P_{\rm tot}$, where $P_{\rm tot}$
is the total (gas plus radiation) pressure.  Both Kerr
($a/M=0.998$) and Schwarzschild models are shown, and a no-torque boundary
condition is assumed at the ISCO, which is why the curves rise upward at
small radii.  The effective optical depth is actually a function of frequency,
and what is depicted here is an average using the Rosseland mean free-free
absorption opacity.  Solid and dashed curves assume $\alpha=0.1$ and 0.01,
respectively.  The lower anomalous stress of the latter implies higher
surface density and higher effective optical depth in both the Kerr and
Schwarzschild cases.
}
\label{taum1}
\end{figure}

Note that the modified blackbody spectrum in equation (\ref{modbb}) has
$I_\nu\le B_\nu(T)$.  This makes sense, as deviations from blackbody imply
that the disk is a thermodynamically less efficient radiator.  Because it
must still radiate the same total flux, the ambient gas temperature must be
hotter than the effective temperature, and the resulting spectrum
therefore extends to higher photon energies.  When integrated over the disk,
the spectral energy distribution must therefore flatten below the canonical
$F_\nu\propto\nu^{1/3}$ as photons are redistributed to higher energies.  These
effects are illustrated in Figure \ref{compthoms}, which compares stellar
atmosphere calculations of relativistic disk spectra with the local blackbody
assumption.

\begin{figure}
\centering
\includegraphics[width=9.truecm]{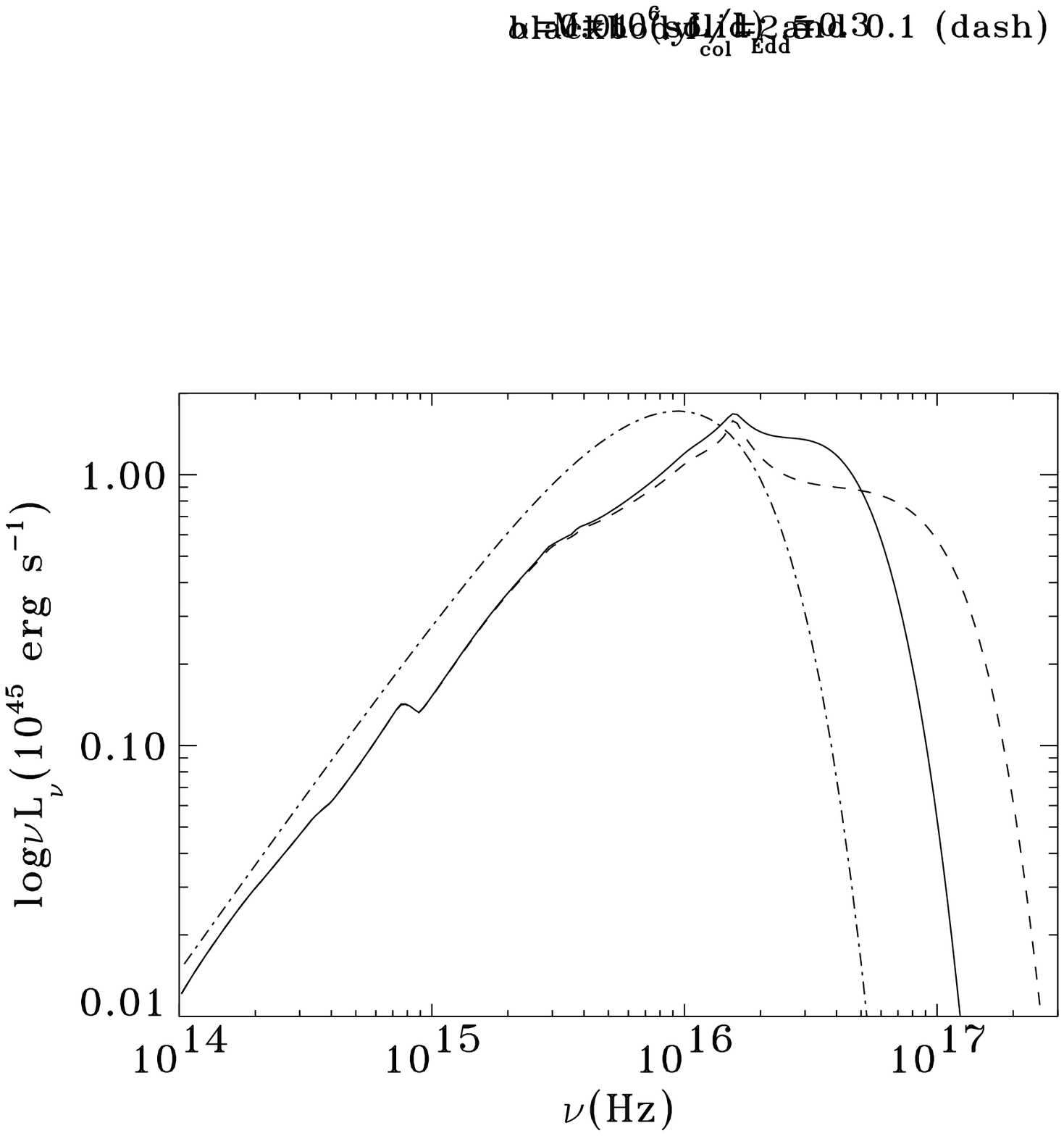}
\caption[]{Spectral energy distributions of relativistic accretion disk models
around a $1.25\times10^8$~M$_\odot$ Kerr ($a/M=0.998$) hole, accreting at
0.25~M$_\odot$~yr$^{-1}$, viewed at 60$^\circ$ from the rotation axis
\cite{hubeny01}.  The dashed curve, which extends to the highest energies,
assumes that electron scattering is coherent (Thomson scattering). The
solid curve allows for the effects of Compton scattering.
Compton downscattering pulls the high energy tail of the spectrum
down, while Compton upscattering increases the spectral luminosity at lower
energies.  The dot-dashed curve is the spectrum that results from assuming
blackbody emission at the local effective temperature at every radius, and
is clearly a very poor approximation.}
\label{compthoms}
\end{figure}

In addition to the effects of Thomson scattering on the spectrum, Compton
scattering can also be important in hot disks.  The net Compton heating
rate of the plasma per unit volume is
\begin{equation}
n_e\sigma_{\rm T}c\left({h\bar\nu\over m_ec^2}-{4kT\over m_ec^2}\right)\left(
{4\pi J\over c}\right),
\label{heatcomp}
\end{equation}
where $J$ is the frequency-integrated mean intensity of the radiation field,
\begin{equation}
J\equiv\int_0^\infty d\nu J_\nu={1\over4\pi}\int_0^\infty d\nu
\oint d\Omega I_\nu,
\end{equation}
and $\bar\nu$ is a specially defined average frequency,
\begin{equation}
\bar\nu\equiv{1\over J}\int_0^\infty d\nu\nu J_\nu
\left(1+{J_\nu c^2\over2h\nu^3}\right).
\label{nubarj}
\end{equation}
(The second term in parentheses in equation [\ref{nubarj}] represents the
effects of stimulated scattering.)  The Compton heating rate can be compared
with the net heating rate from thermal absorption and emission.  Assuming
LTE, this is
\begin{equation}
c\int_0^\infty d\nu\chi_\nu^{\rm th}\left({4\pi \over c}\right)(J_\nu-B_\nu).
\label{heatth}
\end{equation}
Defining the Thomson opacity as
\begin{equation}
\kappa_{\rm T}\equiv {n_e\sigma_{\rm T}\over\rho}
\end{equation}
and the Planck mean opacity as
\begin{equation}
\kappa_{\rm P}\equiv{1\over\rho aT^4}\int_0^\infty d\nu\chi_\nu^{\rm th}\left(
{4\pi B_\nu\over c}\right),
\end{equation}
we see on comparing equations (\ref{heatcomp}) and (\ref{heatth}) that
Compton scattering will dominate the thermal coupling between gas and
radiation if
\begin{equation}
\kappa_{\rm T}\left({4kT\over m_ec^2}\right)\gta\kappa_{\rm P}.
\end{equation}
If, in addition, the effective Compton $y$-parameter,
\begin{equation}
y_{\rm eff}={4kT\over m_ec^2}{\rm max}\left(\tau_{\rm T},\tau_{\rm T}^2\right),
\label{yeff}
\end{equation}
exceeds unity, at the particular frequency of interest, then Compton scattering
will modify the spectrum, either by downscattering if $h\nu>4kT$, or
upscattering if $h\nu<4kT$.  (In equation [\ref{yeff}], $\tau_{\rm T}$ is
evaluated at the $\tau_{\rm eff}=1$ surface.)  Figure \ref{compthoms}
illustrates these effects on the spectrum.

Numerous authors have attempted detailed calculations of the spectral
energy distributions emerging from accretion disks around black holes using
stellar atmosphere modeling at every radius of the disk (see e.g. \cite{sm89,
ln89,rfm92,st93,sha94,st95a,d96,hh97,sk98,hh98,habk00,hubeny01} for models
in the AGN context).  Many of these models fully incorporate general
relativistic effects on the disk structure, and also include
relativistic Doppler shifts, gravitational redshifts, and gravitational
bending of light rays.  These latter effects play an important role in
smearing out atomic absorption and/or emission features in the spectrum,
as illustrated in Figure \ref{cosi99}.  Many models also incorporate
sophisticated treatments of the radiative transfer, Compton scattering, and
non-LTE effects in the atomic level populations of numerous elements and ions.

\begin{figure}
\centering
\includegraphics[width=9.truecm]{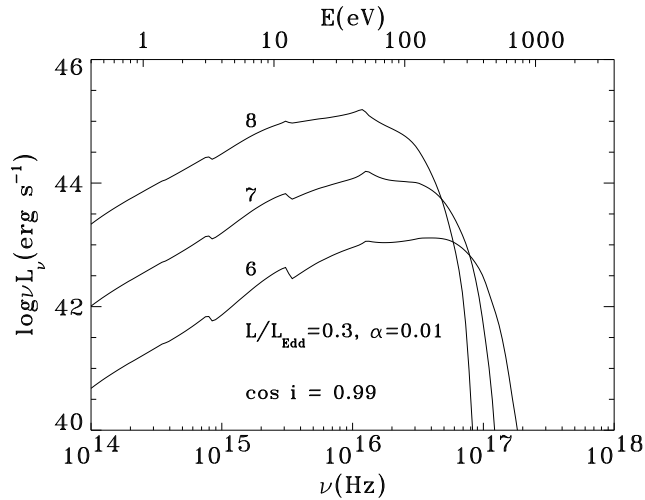}
\caption[]{Spectral energy distributions of accretion disk models around Kerr
($a/M=0.998$) black holes at fixed Eddington ratio viewed nearly face on,
for various black hole masses: $10^8$ (top), $10^7$ (middle), and
$10^6$~M$_\odot$ (bottom) \cite{hubeny01}.  Note the increased relativistic
smearing of the Balmer and Lyman edges of hydrogen for higher black hole mass.
This is because the temperature of the disk is lower at higher mass, so these
features are formed closer to the black hole.}
\label{cosi99}
\end{figure}

As sophisticated as these models are, however, they still rest on variants
of the Shakura-Sunyaev anomalous stress prescription, and are subject to
all the uncertainties we discussed above in section 2.  Moreover, just as
a stellar model must specify a distribution of nuclear energy generation,
a model of the vertical structure of an accretion disk must specify the
distribution of turbulent heating.  The presence of turbulence in the disk also
begs the question as to whether heat is transported radiatively, or whether
bulk transport in the turbulence itself also plays a role.  As we discuss
below in section 5, standard assumptions often used to model the inner,
radiation pressure dominated regions lead to equilibria that are convectively
unstable \cite{bis77}, although it is far from clear how that convection would
manifest itself in a flow that already requires turbulent angular momentum
transport.

Perhaps the most serious question plaguing current Shakura-Sunyaev based
spectral models is the partitioning of the dissipation within the optically
thick and thin regions of the disk.  There is widespread empirical evidence
for the existence of Comptonizing hot plasma in both black hole X-ray binaries
and AGN, and one possibility is that this plasma exists as a magnetized
corona above the disk photosphere.  After all, the turbulent convection zone
of our own sun generates a hot corona, so why shouldn't a turbulent accretion
disk have one as well?  (This is one version of including non-radiative energy
in $F^-$.)  Models have even been considered in which most of
the turbulent dissipation is assumed to occur in the corona, and not in
the disk interior
\cite{hm91}!  If a substantial fraction of the accretion power is in fact
dissipated outside the disk, then there are numerous consequences \cite{sz94}:
the disk becomes geometrically thinner and denser, more effectively optically
thick, and more gas pressure dominated.

Accretion disk physics is uncertain enough, but the physics of the hot plasma
that must exist in real sources is even more insecure.
Whether or not the hard X-rays observed in black hole X-ray binaries or AGN
are generated in a disk corona or elsewhere in the flow (e.g. an inner ADAF
that we discussed at the end of section 2), some of these
X-rays will illuminate the disk from the outside, and thereby be reprocessed.

One of the most interesting X-ray reprocessing features is fluorescent iron
K$\alpha$ line emission.
The great excitement surrounding this line is that in at least
three sources (MCG-6-30-15 \cite{tan95,wilms01,fab02}, Mrk~766 \cite{mas02},
and NGC~3516 \cite{nan99,tur02}), it is clearly observed to be relativistically
broadened.  The resulting line profile is a convolution of the spatial
emissivity profile of the line with relativistic Doppler shifts,
gravitational redshifts, and gravitational light bending \cite{laor91}.
In other words,
the line profile gives us the unprecedented opportunity to map out the
equatorial test-particle orbit structure in a black hole spacetime, assuming
the emitting material (the disk) is geometrically thin.

The K$\alpha$ line is produced because
hard X-ray photons with energies above the K-edge
of iron can knock inner shell electrons out of the atom.  The resulting
ion is born in an excited state and can return to the ground state by having
an L-shell electron drop down to the vacancy in the K-shell, emitting either
a K$\alpha$ line photon or one or more Auger electrons.  The strength of the
resulting emission line is sensitive to the ionization state of iron in
the outer layers of the disk.  At low ionization levels, where both the K
and L shells of most iron ions have no vacancies, the material is transparent
to K$\alpha$ photons which then escape more or less freely from the disk once
they have been created by X-ray fluorescence.  Once iron is sufficiently
ionized that vacancies in the L-shell start to appear (FeXVIII to FeXXIV),
then a K$\alpha$ photon can resonantly scatter off the iron atoms, enhancing
the probability of its destruction through the Auger process.  The emerging
line strength is therefore reduced.  This continues until iron is sufficiently
ionized that only FeXXV and higher is present.  The K$\alpha$ line photons
still resonantly scatter with such ions, but the Auger effect cannot happen
because there are no L-shell electrons to be ejected.
At very high ionization when iron is fully stripped and the recombination
rate is small because of high temperatures, no photoionization can occur and
the emission line disappears.  For an excellent
review of the physics of the iron K$\alpha$ line, see \cite{fab00}.

The energy powering the iron K$\alpha$ line emission comes from photoelectric
absorption of X-rays above the iron K-edge.  Such photoelectric absorption by
iron and other heavy elements present in the disk is in fact an efficient
means of converting illuminating X-ray power below $\sim10$~keV into heat
in the outermost layers of the disk \cite{guil88,lw88}.  Figure \ref{pex}
shows the albedo of an illuminated slab of material as a function of incident
photon energy \cite{mz95}.  Below $\sim10$~keV, most of the incident X-rays
are absorbed due to photoionization of heavy elements, and the level of that
absorption depends on the ionization state of the disk.  Above $\sim10$~keV,
the albedo quickly rises to near unity as the incident X-rays are simply
reflected back by Thomson scattering.  This is widely believed to explain
the characteristic upturn of the X-ray spectrum that is observed at these
energies in both AGN and black hole X-ray binaries \cite{np94,zls99}, and
is strong evidence for the presence of relatively cold material in these
sources.  (Note that especially in the case of AGN, however, this cold material
need not be the disk, but instead could be gas much further out from the
source.)  At higher energies the albedo is again reduced to well below unity,
independent of the ionization state of the slab, because X-rays are Compton
downscattered and therefore lose energy.

\begin{figure}
\centering
\includegraphics[width=9.truecm]{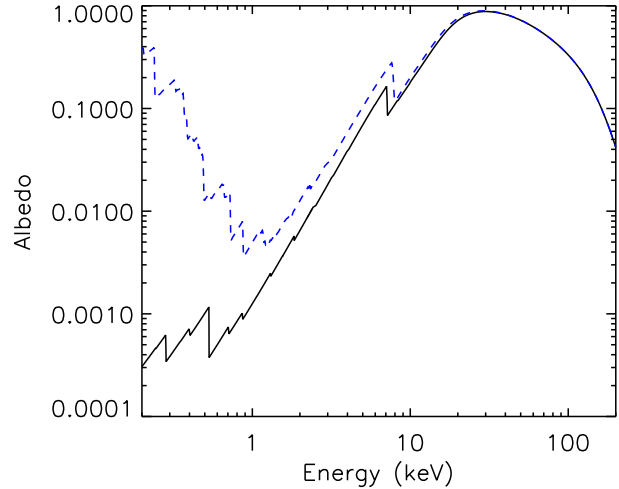}
\caption[]{Energy-dependent albedo of an X-ray illuminated, cold slab with
solar abundances, viewed at twenty degrees from the slab normal.  The solid
curve assumes the slab
is neutral, apart from hydrogen and helium which are fully ionized.  The
dashed curve is for an ionized slab, with ionization parameter
$\xi=100$.  In both cases the illumination was assumed to be by a
power-law spectrum with photon index -2 and exponential cutoff energy of
100~keV.
(This figure was kindly generated by Shane Davis, courtesy of the {\sc XSPEC}
routines {\sc pexrav} and {\sc pexriv} \cite{mz95}.)}
\label{pex}
\end{figure}

Just as the thermal emission spectra from accretion disks can depend
sensitively on the details of the vertical disk structure, X-ray reprocessing
and reflection features also depend on the assumptions one makes in describing
the physics of the illuminated slab.  Models that incorporate hydrostatic
equilibrium produce surface temperature profiles containing
a number of sharp discontinuities due to transitions between different regions
of thermally stable material \cite{nkk00,brf01}, and the resulting
reflection/fluorescent spectra are altered by this fact.

In addition to the
very uncertain geometry of the illumination itself, one wonders whether even
this treatment of the physics is enough.  The bottom line of this section
is that many aspects of our models of observed accreting black hole sources
are fundamentally limited by uncertainties in the basic physics of the
accretion flow itself.  For the rest of these lectures we will switch gears
and address this physics itself, explaining how to go beyond Shakura-Sunyaev
type models.

\section{The Physics of Angular Momentum Transport}

A system with fixed total angular momentum in thermal equilibrium must be in a
state of rigid body rotation \cite{landau80}.  This fact lies at the very
heart of the way accretion disks work, because it shows that differential
rotation is a source of thermodynamic free energy in the flow.  As discussed
long ago by Lynden-Bell \& Pringle \cite{lbp74}, one can always reduce
the total rotational kinetic energy of an accretion disk by fixing its
mass distribution and redistributing the angular momentum so that it is
uniformly rotating.  However, such a state is generally incompatible with
mechanical equilibrium in the gravitational field of the central object, which
requires that the angular velocity decrease outward.  In this case,
pairs of orbiting fluid elements moving with different angular velocities
can lower their mechanical energy by shifting angular momentum from the
element with higher angular velocity to the element with lower angular
velocity \cite{lbp74}.  This transports angular momentum {\it outward}
through the disk, causing material to lose rotational support and flow
in to regions of greater binding energy.  This general behavior is simply
the action of the second law of thermodynamics, and is entirely analogous
to the one-way flow of heat from regions of high temperature to low
temperature.  Ordinary microscopic viscosity acts to transport angular
momentum in exactly this fashion, but this is not sufficient for astrophysical
accretion disks.

Some form of non-microscopic, ``anomalous'' angular momentum transport
mechanism or mechanisms must therefore be at work, and it is important
to recognize that it need not be the same in all accretion flows in the
universe.  Inherent in the Shakura-Sunyaev ``guess'' as to the form of
the anomalous stress $\tau_{r\phi}$ is the existence of some form of
turbulence in the flow:  ideally a form of turbulence that is generated
by the differential rotation itself.  Turbulence is often, though not always,
generated when a laminar flow is dynamically unstable to perturbations.  These
perturbations may be infinitesimal, so that the slightest touch of a feather
will produce instability.  The initial growth of the instability may then
be understood from linearized equations about the equilibrium flow, which
is then said to be linearly unstable.  Linear equations are {\it much}
easier to solve than nonlinear equations, and much of the theoretical work
has therefore been devoted to linear instabilities.  Understanding the
response of the flow to nonlinear, finite amplitude perturbations is a lot
harder, but there do exist laminar flows that are nonlinearly unstable even
while being linearly stable.  Of course to theoretically understand how a
flow really changes in response to the development of even a linear instability
requires a treatment of the nonlinear equations, which in practice is usually
accomplished through numerical simulation.

The simplest way to examine the linear stability of a fluid is to
infinitesimally perturb a fluid element locally and examine whether the
forces exerted on the element act to return it to its original position.
Stability criteria derived in this
way are necessary, but not sufficient because it is also possible for the
global structure of the flow to produce instabilities.  Unfortunately,
analyzing the global stability of an accretion flow is technically much
more difficult, even within linear theory.  One must solve a global
eigenvalue problem for the linear
wave modes of the system, fully accounting for the boundary conditions on
the flow.  In order to derive a stability criterion, one must also
ensure that the modes examined form a complete set so that any
initial perturbation data can be represented in terms of them.

Global stability analyses generally recover the results of local stability
analyses in the limit of very short wavelength perturbations (often called
the WKB limit in the literature).
If global unstable wave modes are found, it may be that changes in the
assumed boundary conditions can stabilize the modes.  This can never be
true of local instabilities, however, which care only about local conditions
in the flow.  In this sense local instabilities are much more robust and
generic than truly global instabilities.

In my discussion of accretion disk instabilities in this section, I will
always neglect the self-gravity of the accreting plasma itself.  Accretion
rates in both AGN and black hole X-ray binaries are
thought to be sufficiently small that the tidal field of the black hole
completely overwhelms the self-gravity of the disk, at least near the
hole where most of the luminosity is generated.  Self-gravity is however
more important further out in the disk in AGN, and may also be important
in hyper-Eddington flows invoked in gamma-ray burst models.


Within ideal hydrodynamics, local linear stability of an axisymmetric rotating
flow is guaranteed if the H{\o}iland criterion is satisfied (e.g.
\cite{tassoul}): 
\begin{equation}
N^2+\kappa^2>0
\label{hoiland1}
\end{equation}
and
\begin{equation}
({\bnabla}P\times{\bf\hat R})\cdot({\bnabla}\ell^2\times{\bnabla}S)>0,
\label{hoiland2}
\end{equation}
where
\begin{equation}
\kappa^2={1\over R^3}{\partial\ell^2\over\partial R}
={1\over R^3}{\bf\hat R}\cdot{\bnabla}\ell^2
\end{equation}
is the square of the epicyclic frequency and
\begin{equation}
N^2 ={1\over\rho}{\bnabla}P\cdot\left({1\over\rho}{\bnabla}\rho
-{1\over\gamma P}{\bnabla}P\right)
={-1\over\rho c_P}{\bnabla}P\cdot{\bnabla}S
\label{brunt}
\end{equation}
is the square of the Brunt-V\"ais\"al\"a frequency.  Other quantities in
these equations are the pressure $P$, the density $\rho$, the entropy per
unit mass $S$, the heat capacity per unit mass at constant pressure $c_P$,
the ratio of heat capacities $\gamma$, the angular momentum per unit mass
$\ell$, the distance from the rotation axis $R$, and the corresponding
cylindrical polar coordinate radial unit vector ${\bf\hat R}$.
(A relativistic generalization of the H{\o}iland criterion can be found in
\cite{seguin}.)

The H{\o}iland criterion is easy to understand in two limits.  For
non-rotating equilibria (e.g. a non-rotating star), the criterion reduces
to the Schwarzschild criterion that the entropy must not increase inward for
stability against convection.  Or, perhaps in more familiar form, the
temperature must not increase inward faster than the adiabatic (constant
entropy!) temperature gradient.  Provided this is true, local fluid elements
will simply oscillate at the local Brunt-V\"ais\"al\"a frequency under
stable buoyancy forces, and these restoring forces are responsible for
supporting gravity waves (g-modes) in stars.  To isolate the effects of
rotation, consider an equilibrium
that has constant entropy everywhere.  Then the H{\o}iland criterion
reduces to the Rayleigh criterion:  the specific angular momentum must
not decrease outward for stability.  Physically, if one perturbs a fluid
element radially outward, it conserves its own specific angular
momentum.  If the ambient specific angular momentum decreases outward,
then the fluid element will be rotating too fast to stay in its new
position, and centrifugal forces will fling it further outward.  Stability
in this case implies that perturbed fluid elements will oscillate back
and forth at the local epicyclic frequency, and organized epicyclic
oscillations form the basis for Rossby waves (r-modes) in rotating stars.
It is no accident that g-mode and r-mode physics appear together in the
H{\o}iland criterion, as these modes are fundamentally intertwined when
both angular momentum and entropy gradients are present.

The H{\o}iland criterion is a huge disappointment for understanding why
turbulence might exist in accretion disks.  The piece of it that is
related to the rotation is connected to specific angular momentum gradients,
and these are strongly stable for rotation profiles that do not differ
too much from Keplerian.  Of course, the H{\o}iland criterion is only
a local stability criterion.  There do exist global, linear hydrodynamic
instabilities in accretion disks.  The one I am most familiar with is
the Papaloizou-Pringle instability \cite{pp84} but, like all global
instabilities, its existence is sensitive to the assumed boundary conditions
\cite{bla87}.  Its nonlinear development also appears to produce large scale
spiral waves rather than local turbulence \cite{haw91}, although such waves
are perfectly capable of producing ``anomalous'' angular momentum transport.
The H{\o}iland criterion also only addresses linear stability.  Hydrodynamic,
planar shear flows can be violently unstable to finite amplitude perturbations
even if they are linearly stable, and this suggests that something similar
might happen in accretion disks.  However, extensive searches for such
instabilities with numerical simulation have consistently failed to find
them in linearly stable flows, except very close to the limit of
marginal stability \cite{bhs96,hbw99}.  It is noteworthy that this
is true even though the same simulation hardware {\it easily} and correctly
finds the instabilities in planar shear flows \cite{bhs96,hbw99}.  As a
point of principle, it is possible that the simulations are somehow
not accessing an unknown destabilizing mechanism (see \cite{long02}
for a recent argument in this direction).  However, one should keep in mind
that there is an important physical difference between linearly stable planar
shear flows and differentially rotating flows: the latter have a strong local
restoring force arising from specific angular momentum gradients \cite{bhs96}.

There is a key piece of physics that is central to the H{\o}iland criterion:
perturbed fluid elements conserve their own entropy and specific angular
momentum.  This is {\it not} true when we include the effects of magnetic
fields, as they fundamentally alter the local dynamics of perturbed fluid
elements.  To order of magnitude, the characteristic frequency associated
with magnetic fields, within the MHD approximation anyway, is simply
$kv_{\rm A}$, where $k$ is the perturbation wavenumber and
$v_{\rm A}=B/(4\pi\rho)^{1/2}$ is the Alfv\'en speed.  Magnetic fields will
alter the character of hydrodynamic r- and g-modes whenever $kv_{\rm A}$
is of order the relevant frequency ($|\kappa|$ or $|N|$) of these modes.
(This is true even when the hydrodynamic modes are unstable, e.g. magnetic
fields can strongly affect hydrodynamic convective instabilities which have
imaginary $N$.)  At first sight one would think that a weak magnetic field
(low Alfv\'en speed) would therefore have no effect.  However, no matter how
weak the magnetic field is, one can always make it important by going to
sufficiently high wavenumbers (short length scales), e.g.
$k\sim|\kappa|/v_{\rm A}$ in the case of r-modes.  Indeed, for
$k\gg|\kappa|/v_{\rm A}$, one expects rotation to be irrelevant as the
modes become magnetically dominated, no matter how weak is the magnetic field.
This all assumes {\it ideal} MHD, in particular that magnetic field lines
are frozen into the fluid.  Resistive effects, which allow the field lines
to ``slip'' with respect to the fluid, become more important on small
scales.  If the field is so weak that it only becomes important below resistive
length scales, then we are back to hydrodynamics on all scales.

The basic physics as to how weak magnetic fields destabilize accretion disks
is most transparent using a mechanical analogy developed by Balbus \& Hawley
\cite{bh92a}.  Consider two identical test particles of mass $m$ moving
together (right on top of each other) in a circular orbit in an external
gravitational potential $\Phi(R)$.  The angular velocity $\Omega$ of the
particles' motion will of course be given by
\begin{equation}
\Omega^2={1\over R}{\partial\Phi\over\partial R}.
\end{equation}
Now suppose we perturb the particles away from their common circular orbit,
but keep them in the same orbital plane.  (Moving them out of the plane
just introduces linearly independent vertical oscillation modes that do nothing
of interest for us here.)  If there were no other forces in the system, the
particles would just execute epicyclic oscillations at frequency $\kappa$
about their original circular orbit.  To make things more interesting, let
us tether the particles together with a spring of spring constant $K$ and
{\it zero} equilibrium length, so that the spring always exerts a tension
which in and of itself tries to pull the particles back together.  Viewed
in a Cartesian frame rotating at the original circular orbit angular velocity,
with origin at the unperturbed particle position, $x$-axis in the radial
direction and $y$-axis in the azimuthal direction, the linearized perturbed
equations of motion of the particles are easy to write down:
\begin{equation}
{{\rm d}^2x_1\over{\rm d}t^2}=2\Omega{{\rm d}y_1\over{\rm d}t}
-R{{\rm d}\Omega^2\over{\rm d}R}x_1+{K\over m}(x_2-x_1),
\label{xspring}
\end{equation}
\begin{equation}
{{\rm d}^2y_1\over{\rm d}t^2}=-2\Omega{{\rm d}x_1\over{\rm d}t}
+{K\over m}(y_2-y_1),
\label{yspring}
\end{equation}
and two similar equations for the accelerations of the second particle which
are identical to these except with the indices 1 and 2 flipped.  The first
terms on the right hand sides of these equations represent the Coriolis
acceleration, the last terms represent the spring accelerations, and the
middle term in equation (\ref{xspring}) is the result of the combined
gravitational and centrifugal acceleration.

It is easy
to find the normal modes of these coupled oscillator equations.  Adding
corresponding equations together, we get
\begin{equation}
{{\rm d}^2\over{\rm d}t^2}(x_1+x_2)=2\Omega{{\rm d}\over{\rm d}t}(y_1+y_2)
-R{{\rm d}\Omega^2\over{\rm d}R}(x_1+x_2),
\end{equation}
\begin{equation}
{{\rm d}^2\over{\rm d}t^2}(y_1+y_2)=-2\Omega{{\rm d}\over{\rm d}t}(x_1+x_2).
\end{equation}
Assuming a time dependence $\propto\exp(-i\omega t)$, we immediately find
that the oscillation frequency $\omega$ is given by
\begin{equation}
\omega^2=R{{\rm d}\Omega^2\over{\rm d}R}+4\Omega^2=\kappa^2,
\end{equation}
which is hardly surprising.  The particles are moving together as one, with
an unstretched spring, executing epicyclic oscillations.  (There is also a
zero frequency mode corresponding to a constant azimuthal displacement of the
particles.)  The other normal modes are far more interesting.
Subtracting corresponding equations, we get
\begin{equation}
{{\rm d}^2\over{\rm d}t^2}(x_1-x_2)=2\Omega{{\rm d}\over{\rm d}t}(y_1-y_2)
-R{{\rm d}\Omega^2\over{\rm d}R}(x_1-x_2)-{2K\over m}(x_1-x_2),
\end{equation}
\begin{equation}
{{\rm d}^2\over{\rm d}t^2}(y_1-y_2)=-2\Omega{{\rm d}\over{\rm d}t}(x_1-x_2)
-{2K\over m}(y_1-y_2).
\end{equation}
Hence the oscillation frequencies for these normal modes satisfy
\begin{equation}
0=\omega^4-\left({4K\over m}+\kappa^2\right)\omega^2+
{2K\over m}\left({2K\over m}+R{{\rm d}\Omega^2\over{\rm d}R}\right).
\label{eqmode2}
\end{equation}
This immediately implies that one of the solutions of this quadratic equation
for $\omega^2$ will be negative, implying an unstable/damped pair of
frequencies, if and only if
\begin{equation}
{2K\over m}+R{{\rm d}\Omega^2\over{\rm d}R}<0.
\end{equation}
All accretion disks around black holes have angular velocities that decrease
outward.  Hence a sufficiently {\it weak} spring {\it always} catalyzes
an instability in the particle motion.  Only if the spring is strong does
it snap the particles back to their equilibrium position.


In it simplest form, the MRI is exactly like this mechanical problem.  Consider
an equilibrium, differentially rotating fluid flow in which there is a
uniform vertical component of magnetic field, as well as an azimuthal
(toroidal) component which,
while nonuniform, is too weak to exert significant stresses on the equilibrium
flow.  Like both r- and g-modes, the MRI involves essentially incompressible
motions, so let us assume for now that the flow can be treated as completely
incompressible (${\bnabla}\cdot{\bf v}=0$).  Also, to keep things simple,
let us neglect buoyancy forces for now by assuming that the equilibrium has
uniform entropy.  Now consider local (i.e. wavelengths much less than
equilibrium length scales) perturbations with wavenumbers purely in the
vertical direction.  The incompressibility condition then immediately implies
that $\delta v_z=0$, i.e. the fluid motions are entirely horizontal.  Similarly,
the condition ${\bnabla}\cdot{\bf B}=0$ implies that there is no perturbation
in the vertical field component, $\delta B_z=0$.

The perturbed flux freezing equations are
\begin{equation}
{\partial\delta B_R\over\partial t}=B_z{\partial\delta v_R\over\partial z}
\label{dbr}
\end{equation}
and
\begin{equation}
{\partial\delta B_\phi\over\partial t}=R\xdrv{\Omega}{R}\delta B_R+
B_z{\partial\delta v_\phi\over\partial z}.
\label{dbphi}
\end{equation}
These equations describe the dynamics of the field lines (the spring!)
themselves.  Vertical gradients of radial and azimuthal velocity stretch
the equilibrium vertical field out into radial and azimuthal field,
respectively.  In addition, the background differential rotation shears out
the perturbed radial field into azimuthal field.

The perturbed fluid momentum equations are
\begin{equation}
{\partial\delta v_R\over\partial t}-2\Omega\delta v_\phi={B_z\over4\pi\rho}
{\partial\delta B_R\over\partial z},
\label{dvr}
\end{equation}
\begin{equation}
{\partial\delta v_\phi\over\partial t}+{\kappa^2\over2\Omega}\delta v_R=
{B_z\over4\pi\rho}{\partial\delta B_\phi\over\partial z},
\label{dvphi}
\end{equation}
and
\begin{equation}
0=-{\partial\delta P\over\partial z}-
{B_\phi\over4\pi\rho}{\partial\delta B_\phi\over\partial z}.
\label{dvz}
\end{equation}
The radial and azimuthal momentum equations (\ref{dvr}) and (\ref{dvphi})
are very similar to the spring equations (\ref{xspring}) and (\ref{yspring}),
the differences being due to the fact that we are now using Eulerian rather
than Lagrangian variables.  In terms of the Lagrangian displacement ${\bxi}$,
the Eulerian velocity perturbations are
\begin{equation}
\delta{\bf v}={\partial{\bxi}\over\partial t}+{\bf v}\cdot{\bnabla}{\bxi}
-{\bxi}\cdot{\bnabla}{\bf v}={\partial\xi_R\over\partial t}\hat{\bf R}+
\left({\partial\xi_\phi\over\partial t}-\xi_RR\xdrv{\Omega}{R}\right)
\hat{\bphi}.
\end{equation}
Note that vertical gradients in the radial and azimuthal field are producing
tension forces in equations (\ref{dvr}) and (\ref{dvphi}) very analogous to
the spring forces considered earlier.
The vertical momentum equation (\ref{dvz}) is rather interesting.
Vertical gradients in
the azimuthal field create vertical magnetic pressure gradients which, for an
incompressible fluid, are immediately balanced by vertical fluid pressure
gradients.  This equation just determines the fluid pressure, and is otherwise
decoupled from the other equations.  Note that in the linear system the vertical
pressure balance equation (\ref{dvz}) becomes trivial if the equilibrium
azimuthal field $B_\phi$ vanishes, and in fact this is the only place where
$B_\phi$ appears.  (We will see in section 5 that radiative diffusion
can affect the MRI through this vertical pressure balance equation, however.)

Assuming a space-time dependence of the perturbations
$\propto\exp[i(kz-\omega t)]$, the four equations (\ref{dbr}), (\ref{dbphi}),
(\ref{dvr}) and (\ref{dvphi}) can be solved to give the dispersion relation
\begin{equation}
0=\omega^4-\left(2kv_{\rm A}+\kappa^2\right)\omega^2+
kv_{\rm A}\left(kv_{\rm A}+R{{\rm d}\Omega^2\over{\rm d}R}\right),
\label{mridisp}
\end{equation}
where $v_{{\rm A}z}=B_z/(4\pi\rho)^{1/2}$ is the Alfv\'en speed corresponding
to the vertical field component.  Equation (\ref{mridisp}) is identical to
eq. (\ref{eqmode2}), if we simply replace the
natural frequency $2K/m$ of the spring with $kv_{{\rm A}z}$!

It is easy to show from eq. (\ref{mridisp}) that the maximum instability
growth rate is
\begin{equation}
|\omega|_{\rm max}={1\over2}R\left|\xdrv{\Omega}{R}\right|,
\label{omegamax}
\end{equation}
and occurs for wavenumbers satisfying
\begin{equation}
(kv_{{\rm A}z})_{|\omega|{\rm max}}=\Omega^2-{\kappa^4\over16\Omega^2}.
\label{eqkcrit}
\end{equation}
Note that this is all in agreement with our statement above that r-mode
physics will be substantially modified when the natural frequency of our
magnetic ``spring'' $kv_{\rm A}$ is of order $\kappa$.

The analysis we have done so far recovers the simplest form of the MRI.  It
turns out that this mode is also an exact solution of the {\it nonlinear}
local MHD equations, although it is itself vulnerable to further
instabilities \cite{gxu94}.  It is often called the ``channel solution'',
as it is characterized by rapid counterstreaming motions in horizontal planes
\cite{hb91,hb92}.

One can of course generalize the analysis to include general axisymmetric
wavenumbers ${\bf k}=k_R\hat{\bf R}+k_z\hat{\bf z}$ and arbitrary equilibrium
field directions.  One can also add buoyancy by introducing entropy gradients.
Most relevant to geometrically thin disks are vertical entropy gradients.  In
this case the MRI dispersion relation can be written as \cite{bh91}
\begin{equation}
0={k^2\over k_z^2}\tilde\omega^4-
\left(\kappa^2+{k_R^2\over k_z^2}N^2\right)\tilde\omega^2-
4\Omega^2({\bf k}\cdot{\bf v}_{\rm A})^2,
\label{bruntmri}
\end{equation}
where $\tilde\omega^2\equiv\omega^2-({\bf k}\cdot{\bf v}_{\rm A})^2$ and
${\bf v}_{\rm A}={\bf B}/(4\pi\rho)^{1/2}$ is the vector Alfv\'en speed
corresponding to the equilibrium field ${\bf B}$.
Note that magnetic fields are therefore directly connected to
both r- and g-modes.  The linear analysis of the MRI can also be extended to
localized, nonaxisymmetric perturbations, which are again subject to
unstable growth \cite{bh92b}.  A fully general relativistic linear analysis
of the MRI has recently been done by Araya-G\'ochez \cite{araya02}.

The general local stability criterion in ideal MHD for rotating, axisymmetric
flows turns out to be the same as the H{\o}iland criterion
[eqs. (\ref{hoiland1}) and (\ref{hoiland2})], but with angular momentum
gradients replaced by gradients in the angular velocity $\Omega$
\cite{balbus95}:
\begin{equation}
N^2+R{\partial\Omega^2\over\partial R}>0
\label{hoilandmhd1}
\end{equation}
and
\begin{equation}
({\bnabla}P\times{\bf\hat R})\cdot({\bnabla}\Omega^2\times{\bnabla}S)>0.
\label{hoilandmhd2}
\end{equation}
Magnetic fields appear to facilitate the existence of instabilities which
are more directly tuned into the thermodynamic sources of free energy in
the flow.  Indeed, Balbus \cite{balbus01} has shown that in situations
where heat conduction exists and is restricted to flow along magnetic field
lines, the entropy gradients in the MHD H{\o}iland criterion are replaced
by temperature gradients!  This situation arises in cases where
charged particles conduct the heat and the collisional mean free
path exceeds the Larmor radius of the particles.  In a non-rotating,
hydrostatic system, the Schwarzschild stability criterion that the entropy
increase upward is replaced by the criterion that the {\it temperature}
increase upward!  While interesting, it should perhaps be remembered
that any temperature or angular velocity gradient constitutes a thermodynamic
source of free energy, but magnetic fields are destabilizing for only
certain directions of these gradients.

The MRI relies on weak magnetic fields.  If the field is too strong at
the wavelength considered, then magnetic tension will overcome the effects
of magnetic torquing and the flow will stabilize.  The stronger the field,
the longer the wavelength required for instability.  This immediately suggests
that laminar accretion disk flow models will be unstable provided an unstable
MRI wavelength $\sim v_{\rm A}/\Omega$ can fit inside the vertical thickness
of the disk.  This corresponds to initial field energy densities that are
less than the thermal pressure.  That this does indeed provide an upper limit
to the field strength in order for the MRI to exist is confirmed by global
linear analyses \cite{cps94,gb94}.

Since its discovery, many numerical simulations have been done by a number
of groups that confirm that the nonlinear development of the MRI in an initially
weakly magnetized medium leads to sustained MHD turbulence in the accretion
flow.  Although the simplest MRI modes are axisymmetric, it is essential that
these simulations be fully three dimensional.  Local axisymmetric simulations
with an initial net poloidal field are inevitably dominated by the channel
solution, which tends to break up in three dimensional simulations.
Axisymmetric simulations with no initial net poloidal field only develop
a transient phase of turbulence that eventually decays away.  This is a
consequence of Cowling's anti-dynamo theorem, which in one version states
that magnetic fields cannot be sustained by fluid motions under conditions
of axisymmetry \cite{mof78}.

Three dimensional nonlinear simulations of MRI turbulence have been done
in a number of geometries, the simplest being the ``shearing box''
\cite{hgb95, hgb96}.  Here
one zooms in on a tiny region of the disk and explores the local nonlinear
behavior of the MRI, neglecting background gradients in the initial equilibrium
except in the radial direction where one has to take into account the all
important differential rotation.  Periodic boundary conditions are adopted
in the vertical and azimuthal directions, and also in the radial direction
in coordinates that shear with the background differential rotation.  (An
excellent discussion of the shearing box can be found in \cite{hgb95}.)
These shearing box simulations show clearly that the nonlinear development
of the MRI is one of fully three dimensional, anisotropic MHD turbulence
that exhibits strongly correlated fluctuations in azimuthal and radial
components of velocity and magnetic field.  Angular
momentum is transported outward by a sum of Reynolds (fluid) and Maxwell
(magnetic) stresses, i.e.
\begin{equation}
\tau_{R\phi}=\rho v_R\delta v_\phi-{B_RB_\phi\over4\pi},
\label{eqreymax}
\end{equation}
where $\delta v_\phi$ is the azimuthal velocity minus the mean background
orbital velocity.  The Maxwell stress generally dominates the Reynolds stress
by factors of 3 to 4 \cite{hgb95}, and their values averaged over the
computational volume both exhibit substantial variability on the orbital
time scale.

Unfortunately, shearing box simulations cannot tell us the magnitude of the
anomalous stress, even in an averaged sense, appropriate for accretion
disks in nature.  The neglect of vertical stratification and buoyancy prevents
the gas pressure from having much dynamical significance in the turbulent
state.  In fact, shearing box simulations carried out with an adiabatic
equation of state have monotonically growing pressure generated by heating
in the turbulence.  However, this increasing pressure does not affect the
average value of $\tau_{R\phi}$ \cite{hgb95}, perhaps because it acts simply
to enforce incompressibility in the turbulence to a greater and greater degree.
Instead of the pressure, the anomalous stress is more closely related to the
overall magnetic energy density in the turbulent state, but this turns out to
depend on the size of the computational box, the initial magnetic field
strength if there is a net mean field in any direction in the box, the
resistivity (numerical or physical) and the size of the artificial viscosity
used in the simulations \cite{hgb95, hgb96}.  The latter two quantities
determine in particular the rate at which magnetic field ``reconnects''
in the simulations.
The initial field topology also affects the level of turbulent
transport.  Simulations with no initial net poloidal field result in a
stress that is roughly an order of magnitude below that obtained when there
is a net initial poloidal field \cite{hgb95, hgb96}.

Simulations have also been done in vertically stratified shearing boxes
\cite{bnst95, shgb96, ms00}.  Such simulations zoom in on a small
vertical slice through the disk, fully taking into account the effects of
vertical buoyancy and a finite scale height of the disk set by vertical
hydrostatic equilibrium.  This latter fact is important, as it connects the
local pressure to the vertical size of the disk.  In simulations that start
out with a weak magnetic field with no net poloidal component, three dimensional
MHD turbulence is again the generic result, at least within the disk interior.
The time and space averaged anomalous stress generated by this turbulence 
turns out to be proportional to the pressure at the midplane of the disk,
in excellent agreement with the Shakura-Sunyaev prescription
(\ref{alphastress}).  The reason for this appears to be that the midplane
pressure sets the vertical scale height.  The thermodynamics of these
simulations is relatively crude: in simulations run with an isothermal
equation of state, so the gas cannot heat or cool, the turbulent stress
saturates to a constant average value.  In simulations run with an adiabatic
equation of state, so the gas continually heats as a result of turbulent
dissipation, the scale height of the disk slowly increases in response to
the heating, and the average turbulent stress also increases in lock step
with the increasing midplane pressure \cite{shgb96}.  In agreement with
the shearing box simulations, the stress is dominated by magnetic fields
rather than velocity fields.

The actual value of the proportionality constant $\alpha$, as defined by
$\tau_{R\phi}$ divided by the midplane pressure, is typically $\sim10^{-2}$,
although different simulations can produce somewhat larger or smaller values
\cite{bnst95, shgb96, ms00}.  It is still not clear that this value is to
be fully trusted, and it should certainly not be assumed to be a dimensionless
constant of nature!  In all the simulations that have been done so far, it
appears that vertical buoyancy of the magnetic field does not play a dominant
role in setting the saturation level of the turbulence, although this
conclusion might be altered once the effects of radiative diffusion are
incorporated (see section 5 below).  Instead, local
balance between magnetic field amplification by the MRI and dissipation
appears to be the most important factor in setting the overall stress level,
just as in the shearing box simulations.
Unfortunately, the dissipation is often largely numerical in nature, and
the actual value of $\alpha$ produced by the simulation can depend
on the grid resolution.  (For example, doubling the grid resolution in
one of the most recent simulations \cite{ms00} increased the value of
$\alpha$ by 1.5.)

One dramatic effect that appears to be a generic feature of all the stratified
shearing box simulations is the formation of a strongly magnetized corona
above the disk \cite{bnst95, shgb96, ms00}.  This has been explored most
extensively by Miller and Stone \cite{ms00} who considered an isothermal
disk (resulting in a vertically Gaussian profile in pressure and density)
within a computational domain that covered plus and minus five scale heights
$H_z$ around the disk midplane.  Figure \ref{fieldlines} depicts the magnetic
field lines in the turbulent state in one of their simulations.  Within
two scale heights of the midplane, MRI turbulence dominates and the field
has a highly chaotic structure.  Above two scale heights, however, the
field becomes much more coherent as the field energy density dominates the
gas pressure in the magnetized corona.  Approximately one quarter of the energy
generated within two scale heights of the midplane is transferred to the
corona in this particular simulation, mostly in the form of Poynting flux
associated with buoyant magnetic field lines.  MRI turbulence therefore appears
capable of self-consistently generating an energetically important corona, in
agreement with the observed fact that the hard X-rays in black hole sources can
carry a substantial fraction of the total accretion power.

\begin{figure}
\centering
\includegraphics[width=9.truecm]{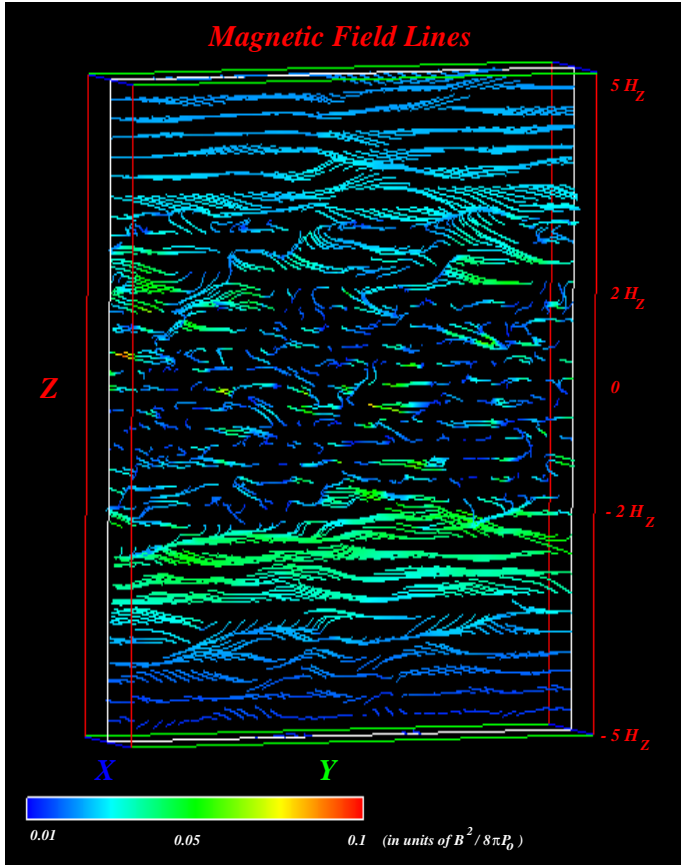}
\caption[]{Magnetic field line structure generated by the MRI in a small
section of an accretion disk, from a simulation by Kristen Miller.  The
figure shows a slice in the Z(vertical)~-~Y (azimuthal) plane of the disk
(X is the radial coordinate).  The color of
the lines represents the strength of the field locally, normalized by the
initial pressure at the disk midplane.  MRI turbulence near the midplane
of the disk (Z$\le 2 H_z$) results in tangled small-scale field structures.
In the lower density
(Z$> 2 H_z$) regions, the field lines become smoother and show large scale
coherence. (This figure was kindly generated by Kristen Miller.  The field
lines in this plot were generated by starting at points in a plane parallel
to the X-Z plane and then integrating the magnetic field line vectors.  This
integration eventually becomes dominated by numerical uncertainties,
particularly in the inner regions of the disk where the field is so chaotic,
at which point the computation of the field line stops for this particular
plot.  Hence the appearance of a nonzero divergence in the field is an
artifact of the way the field lines were computed: ${\bnabla\cdot B}=0$ is
satisfied in the simulations.)}
\label{fieldlines}
\end{figure}


Stratified shearing box simulations have also been done with an initial
net poloidal field, but the results of the most recent such simulations
produce a catastrophic change to a magnetically dominated disk structure
that is inconsistent with the shearing box assumptions \cite{ms00}.
Large scale, ordered magnetic fields that might help generate an MHD outflow
or jet also cannot be formed within the confines of a narrow shearing box.
Global simulations of the entire disk are therefore of great interest, and
some have now been completed.  Examples can be found in
\cite{haw00,mac00,ar01,sh01}.  Such simulations are quite challenging,
and have their own problems when it comes to making inferences about how
real flows in nature work.  Because they attempt to simulate the entire
disk, they are necessarily limited in spatial resolution and total duration
in time.  They also tend to suffer from transient effects because of the
presence of an outer radial boundary condition or an imposed finite radial
extent of the disk itself.  Nevertheless, these simulations have generated
some rather interesting results.

First, time and space-averaged values of the Shakura-Sunyaev alpha stress
parameter range from $\sim0.01$ to $\sim0.2$, consistent with results obtained
from the local simulations.  The reader should once again bear in mind all the
caveats, however.  Second, strongly magnetized coronal regions are generally
formed outside the disk.  Third, outflows are generated, although it is not yet
clear whether a disk that starts out with a weak magnetic field and develops
MRI turbulence can generate a globally ordered field in the corona that can
collimate and accelerate jets.  There are recent intriguing results
where rotating conical outflows that are externally confined by magnetic
pressure are produced near the rotation axis \cite{hbs01,hb02}, but whether
such outflows can produce the apparently self-collimating jets we observe
on large scales in nature is far from clear.  Fourth, significant stresses
are exerted on the flow near the ISCO, a point that I mentioned in section 2
and to which I will return shortly.

Before doing that, I would like to first address the issue as to whether
and how far all this MRI physics can be used to justify the standard accretion
disk equations that we wrote down in section 2.  This has been carefully
examined in the literature by Balbus and collaborators \cite{bgh94,bp99}
in the context of geometrically thin accretion disks, and
I have made an attempt here to generalize their approach to include the effects
of radial pressure support and advection.  Our objective here is to see how
far we can sweep all the MHD physics into a simple anomalous stress
prescription.

Using cylindrical polar coordinates $(R,\phi,z)$, the standard MHD equations
can be manipulated to express local conservation laws in mass,
\begin{equation}
{\partial\rho\over\partial t}+{\bnabla\cdot}(\rho{\bf v})=0,
\label{eqcont}
\end{equation}
radial momentum,
\begin{equation}
{\partial\over\partial t}(\rho v_R)+{\bnabla\cdot}\left(\rho v_R{\bf v}-
{B_R\over4\pi}{\bf B}\right)=-{\partial\over\partial R}\left(P+{B^2\over8\pi}
\right)+\rho\left({v_\phi^2\over R}-{\partial\Phi\over\partial R}\right)
-{B_\phi^2\over4\pi R},
\label{eqradmom}
\end{equation}
angular momentum,
\begin{equation}
{\partial\over\partial t}(R\rho v_\phi)+{\bnabla\cdot}\left[R\rho v_\phi{\bf v}
-{RB_\phi\over4\pi}{\bf B}_{\rm p}+R\left(P+{B_\phi^2\over8\pi}\right)
\hat{\bphi}\right]=0,
\end{equation}
and energy,
\begin{eqnarray}
&&{\partial\over\partial t}\left({1\over2}\rho v^2+\rho\Phi+\rho U+
{B^2\over8\pi}\right)+\nonumber\\
&&{\bnabla\cdot}\left[\left({1\over2}\rho v^2+\rho\Phi+
\rho U+P\right){\bf v}+{1\over4\pi}{\bf B}\times({\bf v}\times{\bf B})+
{\bf F}\right]=0.
\label{eqenergy}
\end{eqnarray}
Here ${\bf B}_{\rm p}=B_R\hat{R}+B_z\hat{z}$ is the poloidal piece of the
magnetic field and ${\bf F}$ is the radiative heat flux.

There is nonaxisymmetric, time-dependent turbulence that is very complicated,
but presumably there is some way of averaging over this turbulence to get
equations describing macroscopic, mean flow properties.  It turns out that
it is possible to (nearly) recover the vertically integrated radial disk
structure equations (\ref{mass})-(\ref{energy}) by such averaging,
{\it provided} we make the following assumptions.  First, the flow velocity
is dominated by rotation, on top of which there are velocity fluctuations with
root mean square amplitudes that are much larger than the mean radial infall
velocity:
\begin{equation}
{\bf v}=R\Omega\hat{\bphi}+{\bf u}\,\,\,\,\,{\rm with}\,\,\,\,\,R\Omega
\gg<u^2>^{1/2}\gg|<u_R>|.
\end{equation}
Also, the turbulence is MHD in character and highly subsonic, with 
\begin{equation}
\left<{B^2\over4\pi\rho}\right>\sim<u^2>\ll<c_s^2>\sim{<P>\over<\rho>}.
\end{equation}
In addition, we assume that there is negligible mean magnetic field, i.e.
$|<{\bf B}>|\ll<B^2>^{1/2}$.  Finally, we assume that negligible fluxes
of mass, angular momentum, and linear momentum leave the upper or lower
disk surfaces, and the only flux of energy leaving these surfaces is radiative
in character, i.e. contained in the radiation flux ${\bf F}$.  We already
know that this last assumption is dangerous, as simulations clearly show
substantial energy transfer into the disk corona.  Provided it is only energy 
and not mass, angular momentum or linear momentum, we will be able to get
away with this by calling ${\bf F}$ an energy flux as I did in section 2.
However, this immediately precludes the possibility of developing large
scale outflows that carry away mass and angular momentum from the disk.

Now, following \cite{bp99}, define a vertically integrated, density weighted,
spatially averaged value of a quantity X at radius $R$ in the disk by
\begin{equation}
<X>\equiv {\hat{A}(\rho X)\over\Sigma}
\equiv{1\over2\pi\Delta R\Sigma}\int_{-H}^H
dz\int_{R-\Delta R/2}^{R+\Delta R/2}dR\int_0^{2\pi}d\phi X\rho,
\end{equation}
where $\hat{A}$ is a vertical integration and averaging operator,
$\Delta R\sim H\ll R$ is a small
radial scale length over which we are averaging (assuming the turbulence causes
variations on size scales smaller than a measure of the disk half-thickness
$H$), and $\Sigma=\hat{A}\rho$ is the surface density.  Applying the operator
$\hat{A}$ to the continuity equation (\ref{eqcont}), and using the boundary
conditions of vanishing mass flux on the disk surface and azimuthal
periodicity, we obtain
\begin{equation}
{\partial\Sigma\over\partial t}+{1\over R}{\partial\over\partial R}(R\Sigma
<u_R>)=0,
\end{equation}
where the radial derivative is now defined by differencing over the
range $(R-\Delta R/2,R+\Delta R/2)$.  Now, by a stationary flow I mean one
in which the average quantities I have constructed are independent of time.
Then $R\Sigma<u_R>$ is a constant, which I may write as
\begin{equation}
\dot M=-2\pi R\Sigma<u_R>.
\label{eqmdot}
\end{equation}
Using the same procedure on the rest of the conservation equations
(\ref{eqradmom})-(\ref{eqenergy}), I obtain
\begin{equation}
0=\bar{\rho}(\Omega^2-\Omega_{\rm K}^2)R-\xdrv{\bar{P}}{R},
\label{eqmeanradmom}
\end{equation}
\begin{equation}
\dot{M}\xdrv{\ell}{R}={{\rm d}\over{\rm d}R}\left[2\pi R^2\left({\Sigma\over
\bar{\rho}}\right)\tau_{R\phi}\right],
\end{equation}
and
\begin{equation}
\dot{M}\left[\xdrv{\bar{U}}{R}+\bar{P}\xDrv{\left({1\over\bar{\rho}}\right)}{R}
\right]=2\pi R^2\left({\Sigma\over\bar{\rho}}\right)\tau_{R\phi}
\xdrv{\Omega}{R}+4\pi RF^-(R),
\label{eqmeanenergy}
\end{equation}
where
\begin{equation}
\tau_{R\phi}\equiv\bar{\rho}\left(<u_Ru_\phi>-\left<{B_RB_\phi\over4\pi\rho}
\right>\right).
\label{meanstress}
\end{equation}
I have been forced to define other averages here, in particular
\begin{equation}
\bar{U}\equiv{<v_RU>\over<u_R>},
\end{equation}
\begin{equation}
{1\over\bar{\rho}}{\partial\bar{P}\over\partial R}\equiv{1\over\Sigma}
{\partial\over\partial R}\left(\left<{P\over\rho}\right>\Sigma\right),
\end{equation}
and
\begin{equation}
{\bar{P}\over\bar{\rho}}\equiv{<Pv_R/\rho>\over<u_R>}.
\end{equation}
The last two equations may not be rigorously consistent, but this is the
usual consequence of attempting to describe a two-dimensional average flow
with one-dimensional, vertically integrated equations.

If I write $\Sigma=2\bar{\rho}H$, then equations
(\ref{eqmdot})-(\ref{eqmeanenergy}) become identical to equations
(\ref{mass})-(\ref{energy}) that we discussed in section 2, with one
important exception.  Note that we recover the true meaning of the
anomalous stress in equation (\ref{meanstress}): the correlated radial and
azimuthal velocity and magnetic field fluctuations that we saw in equation
(\ref{eqreymax})!  We have (almost) succeeded in hiding the underlying
MHD physics in the average flow by placing it entirely inside the anomalous
stress.

The only problem is that I was unable to recover the left hand side of the
radial momentum equation (\ref{radmomentum}) in equation (\ref{eqmeanradmom}),
i.e. the term $\rho v{\rm d}v/{\rm d}R$.  The reason is that I chose to
neglect the magnetic terms in the original exact equation (\ref{eqradmom}).
I am justified in doing this, and still retaining the radial pressure gradient
term, because the turbulence is subsonic.  The same
reasoning, however, forces me to neglect the poloidal velocity terms in this
equation.  If I want to retain them, then I must also retain the magnetic
terms.

In hydrodynamic models of accretion disks with radial advection, the
$\rho v{\rm d}v/{\rm d}R$ term only becomes important near the sonic point,
but it is precisely here that the assumptions underlying the averaging process
giving rise to equations (\ref{eqmdot})-(\ref{eqmeanenergy}) fail utterly.
The poloidal velocities are transonic and the character of the underlying
turbulence {\it must} change.  Magnetic fields are likely to be important
here as they are stretched out and sheared by the large inflow velocities,
and we cannot neglect them.  It is this physics that is the source of
inner torques exerted on the disk across the ISCO \cite{krolik99b,gammie99}.
Rather than localized turbulence over which we can average, we have torques
that are generated by more coherent magnetic field lines formed by flux
freezing in the rapidly infalling fluid.  Averages cannot be performed here
because the time and space dependence of the ``turbulence'' is comparable
to that of the macroscopic flow itself.

Numerical simulations that probe the nature of these {\it non}-MRI inner
torques have been done by a number of authors \cite{haw00,hk01,arc01}.  The
most recent such simulation, which has the highest resolution, confirms the
existence of significant stresses in the plunging region inside the ISCO,
with matter having ten percent more binding energy and ten percent less
angular momentum \cite{hk02}.  The Shakura-Sunyaev $\alpha$ parameter rises
to huge values in the plunging region, but this
is a mis-application of that prescription because it is due merely to
the drop in pressure in this region of the flow.  The flow is highly
time-dependent and cannot be characterized by stationary models flowing
through critical points of ordinary differential equations.
As always, there are still caveats in these simulations and further
work is needed.  In particular, all the simulations that have
been done so far lack a full general relativistic treatment.  Performing
a simulation in a Kerr spacetime in which the black hole can exert magnetic
torques on the accretion disk, thereby supplementing accretion power with
spin power, is likely to be a profitable research direction!

I have focussed in this section on angular momentum transport in black
hole accretion flows, but the existence of turbulence also begs the question
as to whether or not there could be substantial anomalous heat transport as
well.  Unfortunately, none of the simulations that I discussed here can
address this question, as they do not include radiation transport.  In
fact, the thermodynamics of the fluid is generally treated very crudely
by simply adopting an adiabatic ($P\propto\rho^{5/3}$) or isothermal
($P\propto\rho$) equation of state.  Dissipation in the turbulence is not
properly accounted for, and energy is generally not conserved.  It turns
out that incorporating radiation transport directly into the physics of
the MHD turbulence leads to very interesting results, and this will be
the topic of the next and final lecture.

\section{The Role of Radiation Magnetohydrodynamics}

Shakura-Sunyaev based models of standard accretion disks around black holes
are always radiation pressure dominated if the disk luminosity is
anywhere near Eddington.  If we take $\tau_{R\phi}=\alpha P$, with $P$ being
the total (gas plus radiation) pressure, and assume that vertical heat transport
proceeds through radiative diffusion, then the ratio of radiation to gas
pressure is roughly
\begin{equation}
{P_{\rm rad}\over P_{\rm gas}}\sim10^5\alpha^{1/4}\eta^{-2}(1-f)^{9/4}
\left({M\over{\rm M}_\odot}\right)^{1/4}\left({L\over{\rm L}_{\rm Edd}}
\right)^2\left({R\over r_{\rm g}}\right)^{-21/8}{\cal I}^2,
\label{prpg1}
\end{equation}
where $f$ is the fraction of local accretion power that is not dissipated
in the disk interior \cite{sz94}.  As I discussed previously, there has
always been uncertainty as to whether the anomalous stress scales more with
radiation pressure or gas pressure, or indeed something else, and this has
led to ambiguity about the thermal and ``viscous'' stability of the
inner parts of black hole accretion flows.  Even if we take
$\tau_{R\phi}=\alpha P_{\rm gas}$, which is thermally and ``viscously''
stable, the ratio of radiation to gas pressure is still large:
\begin{equation}
{P_{\rm rad}\over P_{\rm gas}}\sim10^4\alpha^{1/5}\eta^{-8/5}(1-f)^{9/5}
\left({M\over{\rm M}_\odot}\right)^{1/5}\left({L\over{\rm L}_{\rm Edd}}
\right)^{8/5}\left({R\over r_{\rm g}}\right)^{-21/10}{\cal I}^{8/5}.
\label{prpg2}
\end{equation}
Note that radiation pressure is most important for models of active galactic
nuclei and quasars that contain supermassive black holes, but it can still be
important for stellar mass black hole X-ray binary models.  As noted in section
3 above, allowing much of the locally dissipated power to vertically escape
from the disk interior in non-radiative fashion ($f\rightarrow 1$) reduces the
importance of radiation pressure \cite{sz94}.

In addition to the issue of secular stability, there are also other
uncertainties in this inner portion of the flow.  Radiation pressure
support against the tidal field of the black hole implies that
\begin{equation}
{n_{\rm e}\sigma_T\over c}F\simeq\Omega_{\rm K}^2\rho z.
\label{radpz}
\end{equation}
The electron number density $n_{\rm e}\propto\rho$, so that the mechanical
equilibrium condition (\ref{radpz}) gives $F\propto z$, or
${\rm d}F/{\rm d}z$ equals a constant.  On the other hand, it has long been
assumed that the vertical distribution of turbulent dissipation per unit mass
is probably constant in the disk interior, i.e. that the power dissipated per
unit volume is proportional to the local density.  Radiative transport of heat
then implies that
\begin{equation}
{\bnabla}\cdot{\bf F}\simeq\xdrv{F}{z}\propto\rho.
\end{equation}
Hence mechanical and thermal equilibrium immediately lead to the conclusion
that the density $\rho$ is independent of height $z$.  Such an equilibrium
is violently unstable to vertical convection \cite{bis77}.  For example,
eliminating the density gradient in equation (\ref{brunt}) leads to
an imaginary Brunt-V\"ais\"al\"a frequency.  More physically, consider an
upward adiabatic perturbation of a fluid element in a constant density
background.  Because the pressure drops with height, the fluid element
expands and becomes less dense than the surrounding constant density
background.  It is therefore buoyant and unstable.

We are therefore lead to the conclusion that heat is transported out of
the disk convectively, not radiatively.  However, this whole argument
is intimately wrapped up in the physics of the MRI.  As noted in the
last section, the MRI modes are tightly coupled to the gravity modes that
are responsible for hydrodynamic convection.  How does convective heat
transport work in the presence of MRI turbulence?  Moreover, the convective
instability arises because of our assumptions about the mass distribution
of dissipation, and this too is determined by the MRI turbulence, so how does
this really work in a vertically stratified, radiation pressure dominated
medium?

It turns out that radiation MHD profoundly affects the MRI, the Parker
instability, and also introduces {\it new} types of dynamical instability
in the flow.  Radiation MHD is just like standard MHD, except that the photon
``gas'' is treated as a separate fluid that couples to the plasma through
absorption, emission, and scattering.
I summarize the basic Newtonian equations of radiation MHD that are useful
in studying the physics of the radiation pressure dominated regions of
high luminosity black hole accretion flows in the appendix.  The reason that
we have to treat the photons in a more careful way than just writing the
total pressure $P_{\rm tot}$ as the sum $P_{\rm gas}+P_{\rm rad}$ is that,
in addition to being effectively thin ($\tau_{\rm eff}<1$) in some cases,
the innermost regions of black hole accretion flows do not have tremendous
total optical depth either.
For $\tau_{R\phi}=\alpha P_{\rm rad}$, the vertical Thomson depth of the
disk midplane is
\begin{equation}
\tau_{\rm T}\sim \alpha^{-1}\eta(1-f)^{-2}
\left({L\over{\rm L}_{\rm Edd}}\right)^{-1}\left({R\over r_{\rm g}}
\right)^{3/2}{\cal I}^{-1}.
\end{equation}
On the other hand, $\tau_{R\phi}=\alpha P_{\rm gas}$ produces a much denser,
more optically thick disk,
\begin{equation}
\tau_{\rm T}\sim 10^4\alpha^{-4/5}\eta^{-3/5}(1-f)^{-1/5}
\left({M\over{\rm M}_\odot}\right)^{1/5}
\left({L\over{\rm L}_{\rm Edd}}\right)^{3/5}\left({R\over r_{\rm g}}
\right)^{-3/5}{\cal I}^{3/5}.
\end{equation}
Even in this case, photons, which are providing the dominant pressure
support, are highly diffusive, especially on small length scales.

Even in a
perfectly electrically conducting fluid, significant radiation pressure
with radiative diffusion alters MHD in two important ways.  First, the
compressibility of the fluid is greatly enhanced on small scales beyond what
would naively be expected based on a very high radiation sound speed.  Even
highly subsonic motions [speeds $\ll(4aT^4/9\rho)^{1/2}$] will be compressible
if they occur on a scale small enough for photons to diffuse during the motion.
Second, temperature fluctuations in the gas tend to be smoothed out by the
radiation field.  These effects have important consequences for the MRI and
the Parker instability, and they also produce wholly new classes of
instabilities.  The nonlinear, turbulent state of the radiation pressure
dominated region of an accretion disk is therefore likely to be very different
from that envisaged by standard accretion disk theory, and the sorts of
modeling we have been discussing up to now may be doing a very poor job of
describing this, energetically most important, region of the flow.

Radiation MHD effects on the Parker instability were considered quite early
in the development of accretion disk theory \cite{sc81}.  Consider an
isolated, straight, horizontal tube of magnetic flux immersed in the vertically
stratified, radiation pressure dominated environment of a black hole
accretion disk.  If the tube has small diameter, radiative diffusion will
tend to smooth out temperature (and therefore radiation pressure) differences
between the interior and exterior of the tube.  The excess magnetic pressure
inside the tube must therefore be balanced largely by a deficit of gas
pressure caused by a lower interior density.  The flux tube therefore has
a tendency to be buoyant.  The standard Parker instability in pure gas disks
implies that flux tubes will be buoyant if $B^2/(8\pi)\gta P$, and this then
suggests that tubes with diameters less than the radiation diffusion length
over an unstable mode growth time will be unstable if
$B^2/(8\pi)\gta P_{\rm gas}$.  That this is in fact true is supported by
numerical simulations of flux tube dynamics in radiation pressure dominated
environments \cite{sc89}.  Specifically, if the flux tubes are strong enough
to self-consistently maintain an anomalous stress
$\tau_{R\phi}=\alpha P_{\rm tot}$, they are extremely buoyant.  If instead
the flux tubes are only as strong to maintain the lower stress
$\tau_{R\phi}=\alpha P_{\rm gas}$, then they can be retained in the medium
for much longer time scales.  This is at least partly due to the fact
that the ambient medium is then much denser and more optically thick, so that
photon diffusion across the flux tubes is much slower.  Unfortunately,
the results of this study are not conclusive, as they fail to take into
account the MRI (whose application to accretion disks was discovered two
years later).  The question remains as to whether the MRI can either generate
the field sufficiently fast to counteract rapid buoyant escape, or directly
modify the buoyant dynamics itself.

Radiation MHD also modifies the behavior of the MRI.  Recall that the MRI
is nearly incompressible.  Even in the nonlinear regime, $B^2/(8\pi)$ is
substantially less than the thermal pressure in the disk interior and the
turbulent motions are highly subsonic.  However, radiative diffusion will
modify this.  From equations (\ref{omegamax}) and (\ref{eqkcrit}), the
characteristic time scale associated with the MRI is the orbital period
and the characteristic length scale is $\sim v_{\rm A}/\Omega$.  The turbulence
will therefore remain incompressible only if the orbital frequency $\Omega$
times the diffusion time $t_{\rm diff}$ over a distance $v_{\rm A}/\Omega$
is substantially greater than unity.  Now,
\begin{equation}
\Omega t_{\rm diff}\sim\Omega\left({\kappa_{\rm T}\rho\over c}\right)
\left({v_{\rm A}^2\over\Omega^2}\right)\sim{\tau_{R\phi}\kappa_{\rm T}\over
c\Omega},
\end{equation}
where the last equality comes from equation (\ref{meanstress}) and the
fact that MRI turbulence is dominated by magnetic rather than Reynolds
stresses.  On the other hand, if heat is transported vertically
by radiative diffusion and the disk is in thermal equilibrium, then
equations (\ref{fcool}) and (\ref{locqbal}) imply
\begin{equation}
{\tau_{R\phi}\kappa_{\rm T}\over c\Omega}\sim
{P_{\rm rad}^2\over\Sigma^2\Omega^2}\sim 1,
\end{equation}
the last equality coming from vertical hydrostatic equilibrium.  Hence
in a radiation pressure dominated disk in which heat is transported radiatively
and the turbulent stress is of order the magnetic stress, then
$\Omega t_{\rm diff}\sim 1$ and the turbulence is therefore expected
to be compressible \cite{tsks02}.  Note that this argument is rather general,
and in particular is completely independent of the Shakura-Sunyaev stress
assumption.

Compressibility affects the MRI in interesting ways.  Consider again the
channel solution that we discussed in section 4.  We noted that the equilibrium
azimuthal field does not affect the linear growth of the channel solution
at all.  In fact this is true for all the linear axisymmetric MRI modes,
whose growth depends on the equilibrium field only through
${\bf k}\cdot{\bf v}_{\rm A}$ from equation (\ref{bruntmri}).  If
there is some equilibrium $B_\phi$ (and if there is not, there soon will
be!), then the channel solution rapidly develops vertical magnetic pressure
gradients.  Normally these do not affect anything as the fluid pressure
is able to balance these through equation (\ref{dvz}).  However, if the
radiation is diffusive on these length scales, then the radiation pressure
is lost.  If the azimuthal field energy density exceeds the true gas pressure,
then not even the gas pressure can balance the vertical magnetic pressure
gradients.  As a result, vertical motions must be excited which take energy
out of the growth of the MRI.  This effect does {\it not} kill the MRI, but
it can dramatically reduce its linear growth rate \cite{bs01}.  If the
azimuthal field energy density exceeds the true gas pressure and if
$B_\phi>B_z$, then the growth rate of the MRI is reduced below the orbital
frequency by the ratio of the gas sound speed to the azimuthal Alfv\'en speed.
This {\it suggests} that growth of magnetic stresses above the gas
pressure may be sluggish, but numerical simulations are really required to 
answer this question.

Such simulations have now been done in {\it non}-stratified shearing boxes,
both in two \cite{tss02} and three \cite{tsks02} dimensions.  Three
dimensional simulations that have a net vertical magnetic flux through the
box develop magnetic stresses that are independent of radiation diffusion
effects - while linear growth of the channel solution may become sluggish,
it still grows!  On the other hand, it appears that radiation diffusion
does diminish the turbulent magnetic stresses that develop when there is
no net vertical magnetic flux through the box.  Unfortunately however,
for the same reasons that we discussed in section 4, shearing box simulations
cannot provide definitive estimates of the level of stress that will exist in
real flows, and vertically stratified and global simulations including
radiative diffusion will need to be done.

Perhaps the most interesting result that has emerged from the recent
shearing box simulations is that radiative diffusion does indeed make
the nonlinear turbulence highly compressible, and large density fluctuations
(by more than a factor 20 \cite{tsks02}!) form and reform in a highly
time-dependent manner.  One important consequence of this fact is that
some of the turbulent energy is {\it directly} dissipated into photon energy
on scales that are completely resolved by the simulation.  This occurs because
photons diffuse out of compressed regions and into rarefied regions,
damping compressive motions in the fluid \cite{ak98}.  Hence not only are
we now getting a handle on the behavior of the anomalous stress with respect
to angular momentum transport in the radiation pressure dominated region,
we are also learning about how the mechanical energy is dissipated into
heat.  Ironically, this is easier to do here than when gas pressure dominates!

There are also entirely new classes of instabilities that are expected to
be present due to radiation pressure and radiative diffusion when the vertical
gravity is included.  One of these is the ``photon bubble'' instability,
first examined by Arons in the context of accreting X-ray pulsars
\cite{arons92}.  Here density fluctuations in the presence of a strong magnetic
field are driven buoyantly unstable by photons diffusing into and heating
(out of and cooling) underdense (overdense) regions.  A local MHD instability
that may have similar physics has also been found to exist in the radiation
pressure dominated region of black hole accretion disks \cite{gam98}.
Fast and slow magnetosonic waves at short wavelengths where radiative
diffusion causes the loss of radiation pressure support of the wave can
also be unstable to periodic driving by the equilibrium vertical forces
(radiation pressure gradients, gas pressure gradients, and gravity), at
least for waves propagating in certain directions \cite{bs01}.  The
characteristic length scale for these instabilities is the gas pressure
scale height (tiny in a radiation pressure dominated medium) and the
growth rate is much larger than the orbital frequency, by a factor
of order the square root of the radiation to gas pressure ratio.  These
instabilities can only exist in Thomson scattering dominated media if magnetic
tension forces help support the wave, which always occurs for wave vectors
that are neither parallel or perpendicular to the equilibrium magnetic
field.  (In some cases, the small absorption
opacity can drive even purely hydrodynamic acoustic waves unstable
\cite{gm96}.)

Simulations have yet to be done that track the nonlinear development of
these instabilities in an accretion disk environment, but Begelman
\cite{beg01} has constructed a one-dimensional, nonlinear periodic shock
train solution that may describe the outcome of the magnetosonic wave
instabilities.   This solution produces a medium that is highly dynamic
and porous to radiation diffusion, and can in fact support fluxes that
are super-Eddington by factors of 10-100 {\it without} driving an outflow
\cite{beg02} (see also \cite{shav98})!

Both the density fluctuations generated in the shearing box simulations
of the MRI and these compressive wave instabilities strongly suggest that
radiative diffusion generates a highly inhomogeneous, very time dependent,
somewhat radiatively porous inner accretion flow.  This speculation needs
to be verified by further work, but if true, there are numerous observational
implications, including possibly super-Eddington fluxes and
modification of local
spectra due to increased thermalization of the photons with the denser
phases of the medium.  The standard reasoning behind the putative existence of
thermal/``viscous" instabilities in this part of the flow will also be
affected.  Suppose for example that a perturbative increase in temperature
increased the porosity of the medium so that photons could escape more
easily.  Such a process could stabilize the disk against a thermal runaway.
It has even been suggested that the radiation pressure dominated region of
the disk may break up into actual discrete dense clumps interpenetrated by
a much hotter, more optically thin medium capable of producing hard X-rays
by Comptonization \cite{kro98}.  Clearly, further numerical simulations
are likely to produce very interesting results!

\section{Conclusions}

After these lectures, particularly the last one in section 5, a student may
be feeling a strong urge to run screaming from the incredibly complex physics
that must be understood before we can truly build realistic models of accreting
black hole sources.  Let me urge the reader that this was not my intent.
The physics is indeed complex (and fascinating), and this essential fact
is the reason why it has proved so hard to move beyond the Shakura-Sunyaev
stress prescription introduced almost three decades ago.  However, this is
not the time to leave the field, but to enter it!  For the first time we
are beginning to investigate the basic physical principles that underly
these flows, and we also have (or at least will have in the very near
future) the simulation hardware that can provide definitive answers to
the concrete questions we are now posing.  The time is now ripe for genuine
and definitive theoretical breakthroughs that will lead to predictive models that
can extract truly useful physical information from observations of accreting
black hole sources.  I encourage you to join in and take part in these
exciting developments.

\acknowledgements
I would like to thank Shane Davis and Kristen Miller for preparing some
of the figures in these lectures, as well as my collaborators (Eric Agol,
Ivan Hubeny, Julian Krolik, and Aristotle Socrates) for their scientific
insights over the years.  I also acknowledge very useful exchanges with
Ian George, John Hawley, Pierre-Yves Longaretti, Patrick Ogle, and Neal Turner.
I am grateful to the organizers of the Les Houches summer school for the
opportunity to participate in this excellent program!
\appendix
\section{The Equations of Radiation Magnetohydrodynamics}

For completeness, I summarize here the basic (Newtonian) equations of
radiation magnetohydrodynamics (e.g. \cite{smn92}) that are relevant to black
hole accretion disk applications.  These are the mass continuity equation,
\begin{equation}
{\partial\rho\over\partial t} +{\bnabla\cdot}(\rho{\bf v})=0,
\label{eqcontrmhd}
\end{equation}
the gas momentum equation,
\begin{equation}
\rho\left({\partial{\bf v}\over\partial t}+{\bf v\cdot\bnabla v}\right)
=-{\bnabla}p+\rho{\bf g}+{1\over4\pi}({\bf\bnabla\times B}){\bf\times B}+
{\kappa_F\rho\over c}{\bf F},
\label{gasmom}
\end{equation}
the total internal energy equation,
\begin{equation}
{\partial (u+E)\over\partial t}+v_j\nabla_j(u+E)+(u+E)\nabla_jv_j
=-p\nabla_jv_j-P_{ij}\nabla_jv_i-\nabla_jF_j,
\label{entot}
\end{equation}
the radiation energy equation,
\begin{eqnarray}
{\partial E\over\partial t}+v_j\nabla_jE+E\nabla_jv_j&=&
-P_{ij}\nabla_jv_i-\nabla_jF_j+\kappa_P\rho caT^4-\kappa_E\rho cE\nonumber\\
&&+\kappa_T\rho c \left({4kT\over m_{\rm e}c^2}-
{h\bar\nu\over m_{\rm e}c^2}\right)E,
\label{raden}
\end{eqnarray}
the radiation momentum equation,
\begin{equation}
{1\over c^2}\left({\partial F_i\over\partial t}+v_j\nabla_jF_i+F_i\nabla_jv_j
\right)=-\nabla_jP_{ij}-{\kappa_F\rho\over c}F_i,
\label{radmom}
\end{equation}
and the flux-freezing equation,
\begin{equation}
{\partial{\bf B}\over\partial t}={\bnabla\times}({\bf v\times B}).
\end{equation}

Equations (\ref{entot})-(\ref{radmom}) have been written in Cartesian tensor
notation, with summation implied over repeated indices.
Although all the equations above are Newtonian equations,
$v/c$ effects have been included
in the radiation terms, and all radiation quantities are defined in the
local fluid rest frame.  The quantities $\rho$, $u$, $p$, and $T$ are
the density, internal energy per unit volume, pressure, and temperature in
the gas, respectively.  They are related by equations of state,
\begin{equation}
u={p\over\gamma-1},
\end{equation}
where $\gamma=5/3$ is the ratio of specific heats for an ionized gas, and
\begin{equation}
p={\rho kT\over\mu},
\end{equation}
where $k$ is Boltzmann's constant and $\mu$ is the mean molecular weight of
the gas particles.  For simplicity, I have neglected the effects of
ionization and recombination in the gas.
The quantity ${\bf v}$ is the gas velocity and ${\bf g}$ is the local
gravitational acceleration presumed to arise exclusively from the central
object (the black hole).

The radiation energy density $E$, the
radiation flux ${\bf F}$, and the radiation pressure tensor $P_{ij}$ are
defined as frequency-integrated angular moments of the specific
intensity in the local fluid rest frame:
\begin{equation}
E=\int_0^\infty d\nu E_\nu=
{1\over c}\int_0^\infty d\nu\oint d\Omega I_\nu({\bf n}),
\end{equation}
\begin{equation}
F_i=\int_0^\infty d\nu F_{\nu i}=
\int_0^\infty d\nu\oint d\Omega n_i I_\nu({\bf n}),
\end{equation}
and
\begin{equation}
P_{ij}={1\over c}\int_0^\infty d\nu\oint d\Omega n_in_j I_\nu({\bf n}).
\end{equation}
The equations can be closed by integrating the radiative transfer equation
directly to solve for the tensor Eddington factor $f_{ij}$, defined by
\begin{equation}
P_{ij}=f_{ij}E.
\end{equation}
The various opacity factors (all in units of cm$^2$/g) are defined in terms
of different frequency averages over the radiation field.  The energy mean,
Planck mean, and flux mean opacities are defined by
\begin{equation}
\kappa_E\equiv{1\over\rho E}\int_0^\infty d\nu
\chi_\nu^{\rm th}(\rho,T)E_\nu,
\label{kappae}
\end{equation}
\begin{equation}
\kappa_P\equiv{4\pi\over\rho acT_{\rm g}^4}\int_0^\infty
d\nu\chi_\nu^{\rm th}(\rho,T)B_\nu(T),
\label{kappap}
\end{equation}
and
\begin{equation}
\kappa_F{\bf F}\equiv{1\over\rho}\int_0^\infty d\nu
[\chi_\nu^{\rm th}(\rho,T)+n_{\rm e}\sigma_T]{\bf F}_\nu,
\label{kappaf}
\end{equation}
respectively.
Here $\chi_\nu^{\rm th}(\rho,T)$ is the thermal absorption
coefficient (in units of cm$^{-1}$) at frequency
$\nu$, $n_{\rm e}$ is the electron number density, and $\sigma_T$ is the
Thomson cross-section.
The Thomson opacity is
\begin{equation}
\kappa_T\equiv{n_e\sigma_T\over\rho}.
\label{kappat}
\end{equation}
Local thermodynamic equilibrium in the gas has been assumed, so that the
thermal source function in the gas is given by the Planck function $B_\nu(T)$
at the gas temperature.

Due to the fact that it dominates the opacity in the inner portions of black
hole accretion disks, I have incorporated the effects of electron scattering
in the equations.  This includes Compton scattering, which can dominate the
thermal coupling between the gas and the radiation in some cases.  I treated
this in the frequency diffusion (Kompaneets) limit by approximating the
radiation field as isotropic in the local rest frame (see e.g. \cite{hubeny01}).
The quantity $\bar\nu$ in the radiation energy equation (\ref{raden}) is
defined as a frequency average over the local rest frame radiation field
in the following way,
\begin{equation}
\bar\nu={1\over E}\int_0^\infty d\nu \nu E_\nu
\left(1+{E_\nu c^3\over8\pi h\nu^3}\right).
\label{nubar}
\end{equation}
(The second term inside the parentheses in this equation represents the effects
of stimulated scattering, which is necessary to include if one wants zero
heat exchange between the gas and radiation when they reach exact thermal
equilibrium.)

For a superb introduction to radiation
hydrodynamics ({\it sans} magnetic fields!), see the book by Mihalas \&
Mihalas \cite{mm84}.

%
%

%
\end{document}